\documentclass[11pt]{article}
\pdfoutput=1
\usepackage[margin=1.0in]{geometry}
\usepackage{setspace}
\usepackage{amsmath,amssymb,graphicx}
\usepackage[pagebackref=true]{hyperref}
\usepackage{slashed}
\usepackage{tikz}
\usepackage{pgfplots}
\usepgfplotslibrary{fillbetween}
\usetikzlibrary{patterns}
\pgfplotsset{compat=1.18} 

\usepackage[T1]{fontenc}

\usepackage[utf8]{inputenc}

\usepackage[greek,english]{babel}

\usepackage{framed}

\usepackage[font=small,labelfont=bf]{caption}
\usepackage{cite}
\usepackage{comment}

\newcommand{\E}{{\mathrm e}}

\newcommand{\ZZ}{{\mathbb{Z}}}

\newcommand{\rmd}{\textrm{d}}

\newcommand{\be}{\begin{equation}}
\newcommand{\ee}{\end{equation}}

\usepackage{xcolor}
\usepackage{bm}

\newcommand{\hratio}{\xi}
\newcommand{\mb}[1]{\mathbf{#1}}
\newcommand{\conft}{\tau}

\title{\bf A Step in Flux to Suppress Axion Isocurvature
}
\author{Priyesh Chakraborty, Junyi Cheng, Matthew Reece, and Zekai Wang \\
{\small Department of Physics, Harvard University, Cambridge, MA, 02138, USA}
}

\begin{document}

\maketitle

\begin{abstract}

The QCD axion in the pre-inflation scenario faces a stringent isocurvature constraint, which requires a relatively low Hubble scale during inflation. If the axion was heavier than the Hubble scale during inflation, its isocurvature is suppressed and the constraint disappears. We point out a novel mechanism for achieving this, relying on the topological nature of a BF-type (monodromy) mass for the axion. Such a mass term has an integer coefficient, so it could naturally have been very large during inflation and exactly zero by the time of the QCD phase transition. This integer can be viewed as a quantized flux, which is discharged in a first-order phase transition that proceeds by the nucleation of charged branes. This mechanism can be embedded in cosmology in several different ways, with tunneling during, at the end of, or after inflation. We provide a detailed case study of the scenario in which the tunneling event occurs during inflation. We also comment briefly on possible UV completions within extra-dimensional gauge theories and string theory. Intriguingly, the phase transition could be accompanied by the emergence of the chiral Standard Model field content from a non-chiral theory during inflation.

\end{abstract}

\tableofcontents

%%%%%%%%%%%%%%%%%%%%%%%%%%%%%%%%%%%%%%%%%%%%		
\section{Introduction}
\label{sec:intro}

\subsection{The axion isocurvature constraint}
\label{subsec:isocurvconstraint}

The QCD axion, proposed as a solution to the Strong CP problem~\cite{Peccei:1977ur, Peccei:1977hh, Wilczek:1977pj, Weinberg:1977ma}, is a compelling dark matter candidate, produced through the misalignment mechanism~\cite{Preskill:1982cy, Dine:1982ah, Abbott:1982af}. From the modern viewpoint, the axion alone does not solve the Strong CP problem, which requires explaining why QCD contributions dominate over other terms in the axion potential. This ``quality problem''~\cite{Georgi:1981pu, Lazarides:1985bj, Casas:1987bw, Kamionkowski:1992mf, Holman:1992us, Barr:1992qq, Ghigna:1992iv} can be solved by invoking additional gauge symmetries that protect an accidental Peccei-Quinn (PQ) symmetry (e.g.,~\cite{Barr:1992qq, Dine:1992vx, Randall:1992ut}), by considering an extra-dimensional axion which is a zero mode of a higher dimensional gauge field (e.g.,~\cite{Witten:1984dg, Choi:2003wr, Conlon:2006tq, Svrcek:2006yi}), or by hybrids of the two mechanisms (e.g.,~\cite{Barr:1985hk, Cheng:2001ys, Cicoli:2013cha, Petrossian-Byrne:2025mto}). There are two qualitatively different possibilities for axion cosmology (see~\cite{Kawasaki:2013ae, Marsh:2015xka, Safdi:2022xkm, OHare:2024nmr} for reviews): with (``pre-inflation'') or without (``post-inflation'') a phase transition after inflation that spontaneously breaks a PQ symmetry and produces axion string defects. In this paper, we focus on the pre-inflation scenario, in which the axion is treated as an independent real scalar field during inflation, not the phase of a complex scalar stabilized at the origin. There are at least two reasons why the pre-inflation scenario may be theoretically preferred. First, it avoids a tension between solving the quality problem and having a realistic cosmology, which afflicts 4d axion models~\cite{Lu:2023ayc}. Second, extra-dimensional axion models, which solve the quality problem efficiently and arise in top-down string theory constructions, are intrinsically of the pre-inflation type because there is no 4d PQ symmetry to break~\cite{Cicoli:2022fzy,Reece:2023czb, Benabou:2023npn,Reece:2025thc}.

The most significant cosmological puzzle in the pre-inflation QCD axion scenario is how to achieve the correct dark matter abundance without overproducing isocurvature perturbations. Scalar fields that are light during inflation have typical fluctuations of order the Hubble scale, while massive scalar fields have exponentially diluted fluctuations. This can be seen from the positive-frequency mode functions of the fields. Specifically, the mode function of a massless scalar in de Sitter space with Hubble scale $H_I$, as a function of conformal time $\tau = -\frac{1}{H_I} \exp(-H_I t)$, is 
\begin{equation}
\label{eq:masslessmodefunction}
    u_k(\tau) = (i - k \tau)\frac{H_I}{\sqrt{2 k^3}}\E^{-i k \tau}.
\end{equation}
On the other hand, the mode function for a field of mass $m > \frac{3}{2}H_I$ in de Sitter space is 
\begin{equation} \label{eq:massivemodefunction}
    u_k(\tau,\mu) = \frac{H_I\sqrt{\pi}}{2} \mathrm{e}^{i \pi/4}\mathrm{e}^{-\pi \mu/2}(-\tau)^{3/2}\mathrm{H}^{(1)}_{i\mu}(-k \tau) \approx \frac{H_I}{\sqrt{2 \mu}}\left(-\frac{k\tau}{2}\right)^{i \mu}\E^{i \mu(1-\log\mu)}(-\tau)^{3/2},
\end{equation}
where $\mu = \sqrt{m^2/H_I^2 - 9/4}$ and in the last step we took the late-time ($\mu > (k \tau)^2$) limit of the Hankel function. Thus we see that the massive mode function decays exponentially at late times, as $(-\tau)^{3/2} \sim \exp(-3H_I t/2)$, while the massless mode function does not.\footnote{For massive fields, particle production is Boltzmann suppressed. Naively the $\mathrm{e}^{-\pi \mu/2}$ factor in the mode function might hint that isocurvature perturbations are as well, but instead this factor is compensated by an $\mathrm{e}^{+\pi \mu/2}$ factor in the large-$\mu$ scaling of the Hankel function. Isocurvature from massive fields is exponentially diluted by the expansion of the universe, but not Boltzmann suppressed. \label{fn:Boltzmann}} 

In the pre-inflation scenario, the axion is a light scalar field during inflation (and is not the inflaton), so it has independent fluctuations that eventually become dark matter isocurvature perturbations. (In the post-inflation scenario, by contrast, the axion is the phase of a massive complex scalar, and its isocurvature perturbations are negligible.) Data from Planck~\cite{Planck:2018jri} requires that dark matter perturbations be dominantly adiabatic. For axion dark matter, the field value between inflation and the QCD phase transition is frozen on superhorizon scales by Hubble friction,  implying that the conserved adiabatic mode has zero perturbations, $\delta \theta_k(x) \propto \frac{\rmd}{\rmd t}{\overline{\theta(x)}} = 0$. Weinberg's theorem on the conservation of adiabatic modes~\cite{Weinberg:2003sw} then implies that such an initial lack of perturbations becomes the standard cold dark matter adiabatic mode, with $\delta \rho_\textsc{DM}/\rho_\textsc{DM} = \frac{3}{4} \delta \rho_\gamma/\rho_\gamma$, once the axion field begins to oscillate in its potential around the time of the QCD phase transition (see~\cite{Axenides:1983hj} for an early discussion). Nonzero initial fluctuations $\delta \theta_k$, on the other hand, behave as isocurvature modes. Pre-inflation axion dark matter can satisfy the empirical upper bound on isocurvature perturbations, but this imposes an upper bound on the Hubble scale during inflation. This is not a {\em problem}, per se, but it is in tension with many specific models of inflation. It is also in tension with a general expectation that the least fine-tuned models of inflation, and those that are most robust against inhomogeneous initial conditions, have a large Hubble scale~\cite{Linde:1985ub, Linde:2014nna, East:2015ggf, Clough:2016ymm, Aurrekoetxea:2019fhr}.

A summary of the quantitative constraints on the scale of inflation and the amount of isocurvature is as follows. The scalar power spectrum is measured by Planck to be\footnote{See the original Planck publication~\cite{Planck:2018jri} for details of uncertainties and fits with different assumptions; the latest ACT data~\cite{ACT:2025tim} favors slightly higher values of $n_s$, which in turn slightly weakens the isocurvature constraint, but quantitatively has only a mild effect on anything we discuss in this paper.}
\begin{equation} \label{eq:scalarpower}
{\cal P}_{{\cal R}{\cal R}}(k) = A_s \left(\frac{k}{k_{\text{pivot}}}\right)^{n_s - 1}, \quad A_s \approx 2.1 \times 10^{-9}, \quad n_s \approx 0.965,
\end{equation}
whereas for uncorrelated isocurvature perturbations the amplitude is constrained to be
\begin{equation} \label{eq:isocurvconstraint}
    \beta_\mathrm{iso} = \frac{{\cal P}_{{\cal I}{\cal I}}}{{\cal P}_{{\cal R}{\cal R}} + {\cal P}_{{\cal I}{\cal I}}} < 0.038.
\end{equation}
For an axion model, we expect
\begin{equation} \label{eq:axionisocurv}
    \beta_\mathrm{iso}^\mathrm{(axion)} = \left(\frac{\Omega_a}{\Omega_c}\right)^2 \frac{1}{A_s} \frac{H_I^2}{\pi^2 \langle \theta_i^2\rangle f_I^2},
\end{equation}
where $\Omega_a$ and $\Omega_c$ are the abundance of axion dark matter and all cold dark matter, $H_I$ is the Hubble scale during inflation (more precisely, evaluated at the time the $k$ mode used to place the constraint exited the horizon), $f_I$ is the axion decay constant during inflation, and $\theta_i$ is the initial value of the axion field. (The notation $\langle \theta_i^2 \rangle$ allows for the case where the initial value is not precisely constant, but includes an average over fluctuations.) The abundance of QCD axion dark matter produced by the misalignment mechanism depends on the temperature evolution of the axion mass $m_a(T)$. This is now under fairly good theoretical control, with lattice calculations confirming the dilute instanton gas scaling behavior (albeit with a larger prefactor)~\cite{Borsanyi:2016ksw}. We adopt the estimate~\cite{ParticleDataGroup:2024cfk}
\begin{equation} \label{eq:axiondmabundance}
    \frac{\Omega_a}{\Omega_c} \approx \gamma_\mathrm{dil} \gamma_\mathrm{anh} \langle \theta_i^2 \rangle \left(\frac{f_a}{9 \times 10^{11}\,\mathrm{GeV}}\right)^{1.165}, 
\end{equation}
where $\gamma_\mathrm{dil}$ is a possible dilution factor arising from late-time entropy production, which is a reasonable expectation in supersymmetric settings from saxion or moduli decays (see, e.g.,~\cite{Steinhardt:1983ia,Lazarides:1987zf,Lazarides:1990xp,Kawasaki:1995vt,Banks:1996ea, Hashimoto:1998ua, Banks:2002sd}), and $\gamma_\mathrm{anh}$ is a correction for anharmonic effects in the potential, most important if the initial $\theta_i$ is close to $\pi$ (see, e.g.,~\cite{Turner:1985si,Lyth:1991ub,Visinelli:2009zm}). For large values of $f_a$, the correct amount of dark matter abundance can be achieved by taking $\theta_i \ll 1$ (often referred to as the ``anthropic axion window''~\cite{Linde:1991km, Hertzberg:2008wr}) or $\gamma_\mathrm{dil} \ll 1$ (though this is limited by the requirement that the universe reheat above the BBN temperature). A third alternative, the stochastic axion scenario~\cite{Graham:2018jyp,Takahashi:2018tdu}, requires an extremely low $H_I$ and is not relevant for our goals in this paper. Combining~\eqref{eq:isocurvconstraint},~\eqref{eq:axionisocurv} and~\eqref{eq:axiondmabundance}, we obtain the following constraints on the scenario in which axions constitute only a fraction of dark matter, and on two scenarios in which they constitute all of dark matter:
\begin{align}
\label{eq:fractionalbound}
    \text{Fractional DM:} \quad  H_I & < \left(\frac{f_I}{\theta_i f_a}\right)\,  \left(\frac{f_a}{10^{12}\,\mathrm{GeV}}\right)^{-0.165}\, 2.5 \times 10^7\,\mathrm{GeV}, \\
\label{eq:anthropicbound}
    \text{Anthropic:} \quad  H_I & < \left(\frac{f_I}{f_a}\right)\,  \left(\frac{f_a}{10^{12}\,\mathrm{GeV}}\right)^{0.4175}\, 2.6 \times 10^7\,\mathrm{GeV}, \\
\label{eq:dilutionbound}    
    \text{Dilution:} \quad  H_I & < \theta_i \, \left(\frac{f_I}{10^{12}\,\mathrm{GeV}}\right)\, 2.8 \times 10^7\,\mathrm{GeV}.
\end{align}
In the first case (``fractional DM'') , we leave $\theta_i$ arbitrary and set $\gamma_\mathrm{dil} = 1$, leaving the axion to constitute only a fraction of the dark matter. This bound is inapplicable in the regime of large $f_a$, where~\eqref{eq:axiondmabundance} predicts too large an abundance of axion dark matter.
In the second case (``anthropic''), we have set $\gamma_\mathrm{dil} = 1$ and tuned $\theta_i$ to achieve the correct dark matter abundance. This bound is inapplicable when $f_a$ is too small to produce the correct dark matter abundance for any $\theta_i$. In the final case  (``dilution''), we have tuned $\gamma_\mathrm{dil}$ to achieve the correct dark matter abundance. In all cases, we have fixed $\gamma_\mathrm{anh} = 1$, an approximation that will break down for sufficiently large $\theta_i$. We have also left open the possibility that $f_I \neq f_a$, though we will not pursue this possibility further in this work.

The absence of observed primordial tensor modes in the CMB, with the latest bound on the tensor-to-scalar ratio $r < 0.036$ from BICEP-Keck~\cite{BICEP:2021xfz}, implies that the Hubble scale during inflation is bounded above by 
\begin{equation} \label{eq:HItensor}
H_I = \pi \sqrt{\frac{A_s r}{2}} M_\mathrm{Pl} \approx 1.0 \times 10^{-4} \sqrt{r} M_\mathrm{Pl} \leq 4.5 \times 10^{13}\,\mathrm{GeV}.
\end{equation}
This is a substantially weaker bound than~\eqref{eq:anthropicbound} and~\eqref{eq:dilutionbound} derived from axion isocurvature. 

We confine our discussion to the standard axion isocurvature constraint, but note that there may be additional model-dependent isocurvature constraints arising from a blue-tilted spectrum~\cite{Kasuya:2009up, Chung:2017uzc} or non-gaussianities induced by isocurvature~\cite{Kawasaki:2008sn,Hikage:2012be} including oscillatory cosmological collider signals~\cite{Chen:2023txq}.

\subsection{Relaxing the axion isocurvature constraint}
\label{subsec:literaturereview}

The tension between a viable pre-inflation axion and high-scale inflation is widely appreciated~\cite{Seckel:1985tj, Lyth:1989pb, Lyth:1992tx, Kawasaki:1997ct, Fox:2004kb, Beltran:2006sq, Kawasaki:2007mb, Hertzberg:2008wr, Mack:2009hv}. Two main ideas have been proposed for relaxing this tension: \begin{enumerate}
\item {\bf Time-dependent mass.} $m_a(t_I) \gtrsim H_I \gg m_a(t_\mathrm{QCD})$. A field that is more massive than the Hubble scale during inflation has exponentially suppressed perturbations, as we saw in~\eqref{eq:massivemodefunction}, and so does not contribute significantly to isocurvature. Specific mechanisms for achieving a time-dependent mass that have been discussed in the literature include making the QCD scale time-dependent and stronger in the early universe~\cite{Dvali:1995ce, Jeong:2013xta, Choi:2015zra, Co:2018phi, Heurtier:2021rko} or breaking PQ symmetry badly in the early universe~\cite{Dine:2004cq, Higaki:2014ooa, Dine:2014gba, Kawasaki:2015lea, Takahashi:2015waa,Kearney:2016vqw, Buen-Abad:2019uoc, Jeong:2022kdr}. (In some cases, these papers discuss the relaxation of the axion to its minimum during inflation; if this minimum can be approximately aligned with the late-time minimum through some symmetry argument, this can decrease the dark matter abundance, in addition to affecting isocurvature.) A general difficulty with this scenario is the implausibly extreme nature of the dynamics needed to smoothly change the axion mass from a large value (say, $10^{14}\,\mathrm{GeV}$) during inflation to a small value (say, $10^{-6}\,\mathrm{eV}$) today, without disturbing the mechanism that solves the quality problem. Another approach is to consider a plasma of hidden-sector monopoles that gave the axion a large mass and suppressed isocurvature, before diluting away~\cite{Kawasaki:2015lpf, Nomura:2015xil, Kawasaki:2017xwt} (building on the results of~\cite{Fischler:1983sc}). 
\item {\bf Time-dependent decay constant.} $f_I \gg f_a$. Because the axion isocurvature fluctuations are proportional to $H_I / (2\pi f_I)$, taking $f_I$ larger can reduce isocurvature, whereas only the late-time value $f_a$ determines the dark matter abundance, as is apparent in~\eqref{eq:anthropicbound} and~\eqref{eq:dilutionbound}. This has been considered in many papers~\cite{Linde:1990yj, Linde:1991km, Kasuya:1996ns, Kasuya:1997td, Folkerts:2013tua, Kawasaki:2013iha, Chun:2014xva, Fairbairn:2014zta, Nakayama:2015pba, Harigaya:2015hha, Kearney:2016vqw, Kobayashi:2016qld, Allali:2022yvx, Graham:2025iwx}. For an extra-dimensional axion, it is plausible that this scenario could be realized by the dynamics of the volume modulus~\cite{Conlon:2022pnx,Chandrashekar:2025nn}.
\end{enumerate}
One could also consider a mix of the two; for instance, if the inflaton is a modulus that controls the QCD coupling constant, it would alter both the mass and decay constant of the axion. A third option, discussed in the literature, is an alternative way to realize $f_a \gtrsim H_I$ theories, by placing them in the post-inflation rather than pre-inflation scenario~\cite{Lyth:1992tw, Asaka:1998ns, Banks:2002sd, Hertzberg:2008wr, Bao:2022hsg}. Because we are concerned with intrinsically pre-inflationary axions, this route is not relevant for us.

\medskip

\begin{figure}[thbp]
    \centering
    \includegraphics{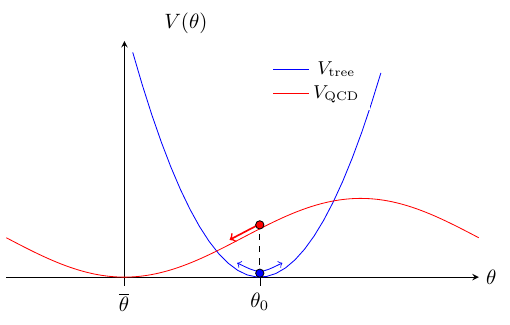}
    \caption{Axion potential schematically: An axion 3-form coupling generates a BF type monodromy potential $V_{\mathrm{tree}}$ (blue). It traps the axion field around its minimum $\theta_0$ and suppresses isocurvature perturbations during inflation. The QCD phase transition turns off the monodromy mass and generates $V_{\mathrm{QCD}}$ (red). The minima of the two potentials are misaligned by an angle $\theta_0 - {\bar \theta}$, which seeds axion dark matter after inflation.}
    \label{fig:misalignment-angle}
\end{figure}

In this paper we explore a novel variation on the first case, a time-dependent mass, in a form that we find much more plausible than past realizations of the idea. The key observation is that there is a well-known way to give an axion a large, tree-level mass, also known as a {\em monodromy mass}, by coupling the axion to a 3-form gauge field~\cite{Silverstein:2008sg}. This mass is of BF type (or, from a dual viewpoint, Stueckelberg type), proportional to an interaction coefficient that takes on only discrete values: an integer, when fields are properly normalized. Because this mass is proportional to an integer, it can naturally be zero. It could also be nonzero and very large. By adding dynamics that can change the value of the integer coupling, we could imagine that a large, nonzero axion mass during inflation exponentially suppresses isocurvature perturbations, but then turns off at late times. By the time of the QCD phase transition, the monodromy mass is precisely zero. The tree-level axion potential is minimized at some value $\theta(x) = \theta_0$, whereas the QCD axion potential is minimized at the value $\theta(x) = \bar \theta$ that solves the Strong CP problem. The values of $\theta_0$ and $\bar \theta$ are not related, as illustrated in Fig.~\ref{fig:misalignment-angle}. Thus, the initial condition $\theta(x) = \theta_0$ can be far from $\bar \theta$ and allow for standard misalignment production of dark matter. The mechanism that made the axion heavy during inflation also has no impact on the axion quality problem in the late-time universe. 

%%%%%%%%%%%%%%%%%%%%%%%%%%%%%%%%%%%%%
%%%%%%%%%%%%%%%%%%%%%%%%%%%%%%%%%%%%%%%%%%%%%%%%

A dynamical integer is one that jumps discontinuously across dynamical domain walls. The cosmology of our scenario, then, requires a first order phase transition from a state where the axion is very heavy to a state where it is light, driven by nucleation of bubbles made of such a domain wall. One route to UV completing such an effective theory is to view the dynamical integer as the flux of a gauge field through extra dimensions. In this case, the domain walls are branes carrying magnetic charge under the extra-dimensional gauge field.

In the remainder of this paper, we explore this idea, working our way from the most general and universal aspects to the most model-dependent. In \S\ref{sec:mechanism}, we explain the underlying mechanism: what a monodromy mass is, and how its coefficient can be dynamical.  In \S\ref{sec:cosmology}, we discuss how the cosmological phase transition from a heavy axion to a light one proceeds. For example, the phase transition could happen during inflation, at the end of inflation, or after inflation, and the possible cosmological signals of these cases differ. In \S\ref{sec:casestudy}, we provide a detailed case study of the scenario in which the first order phase transition happens during inflation but only changes the Hubble expansion rate by a small amount. For this case, we quantify isocurvature constraints on the scenario. In \S\ref{sec:exclusionHf}, we present summary plots of the allowed regions in the $(H_I, f_a)$ plane in standard scenarios and applying our mechanism with different assumptions. In \S\ref{sec:models}, we explore how the mechanism could be embedded in more complete models. We elaborate on the extra-dimensional picture of the origin of the axion monodromy mass, and comment on how the necessary ingredients might be embedded in a string theory UV completion. In \S\ref{sec:outlook}, we offer concluding remarks and discuss a number of questions raised by this work, which would be interesting to explore in the future.

%%%%%%%%%%%%%%%%%%%%%%%%%%%%%%%%%%%%%%%%%%%%		
\section{The Mechanism: First-Order Phase Transition in Axion Mass}
\label{sec:mechanism}

In this section, we present the universal features of our proposed mechanism, leaving a more detailed look at its embedding in cosmology and particle physics for later sections. 

\subsection{Review: axion monodromy mass}
\label{subsec:monodromy}

Our mechanism assumes that during inflation, the axion field $\theta(x)$ has a large tree-level {\em monodromy} mass. In this subsection, we review the origin of such a mass. It arises from a BF-type interaction (i.e., a two-field Chern-Simons interaction) with a 4-form gauge field strength,
\begin{equation} \label{eq:thetaF4}
\int \frac{n}{2\pi} (\theta - \theta_0) F_4,
\end{equation}
where $F_4 = \rmd A_3$ is the field strength of a 3-form gauge field and $n \in \ZZ$ is a Chern-Simons level. In the simplest version of this scenario, we take the axion and the 3-form gauge field to have quadratic kinetic terms,
\begin{equation}
\int \left(-\frac{1}{2} f^2 |\rmd\theta|^2 - \frac{1}{2 e_A^2} |F_4|^2 \right),
\end{equation}
where we adopt the notation $|\omega|^2 \equiv \omega \wedge \star \omega$, the fields are normalized so that $\rmd \theta$ and $F_4$ have fluxes valued in $2\pi \mathbb{Z}$, $f$ is the axion decay constant and $e_A$ (which has units of energy squared) is the gauge coupling of $A_3$. A 3-form gauge field in four spacetime dimensions has no propagating degrees of freedom, so there is only a single massive mode corresponding to the one propagating degree of freedom in $\theta(x)$.

\begin{figure}[htbp]
    \centering
    \includegraphics{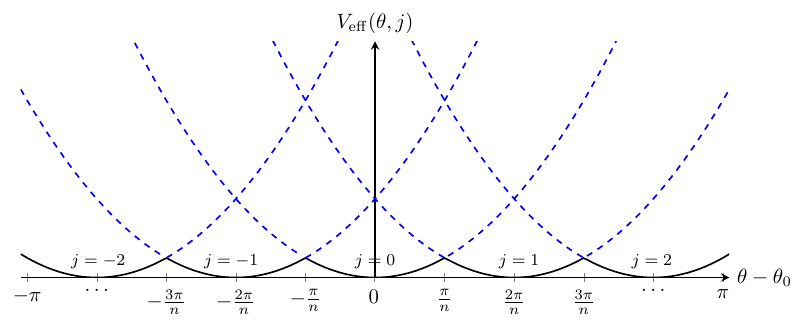}
    \caption{Axion monodromy potential in effective theory: $V_{\mathrm{eff}}(\theta,j)=\frac{1}{2}m_{\theta}^2f^2\left((\theta-\theta_0) - 2\pi j/n\right)^2$. The potential exhibits monodromy when $\theta \to \theta + 2\pi/n$, $j \to j+1$. Dynamical $A$-branes interpolate between branches with different $j$.}
    \label{fig:monodromy-potential}
\end{figure}

In this model, the effective potential for the axion has an infinite set of quadratic branches labeled by an integer $j$, and can be derived by a standard electric-magnetic duality procedure (see, e.g.,~\cite{Kaloper:2008fb, Kaloper:2011jz}):
\begin{equation}
V_\mathrm{eff}(\theta, j) = \frac{1}{2} m_\theta^2 f^2 \left(\theta - \theta_0 - \frac{2\pi j}{n}\right)^2, \, \quad m_\theta = \frac{n e_A}{2\pi f}.
\label{eq:Axion-Effective-Potential}
\end{equation}
This potential is illustrated in Figure~\ref{fig:monodromy-potential}. 

%%%%%%%%%%%%%%%%%%%%%%%%%%%%%%%%%%%%%%%%%%%%
%%%%%%%%%%%%%%%%%%%%%%%%%%%%%%%%%%%%%%%%%

The integer $j$ is the electric flux that is Hodge dual to $F_4$; more precisely, we have
\begin{equation} \label{eq:jF4relation}
    j = \frac{\delta {\cal L}}{\delta F_4} = -\frac{1}{2e_A^2} {\star F_4} + \frac{n}{2\pi} (\theta - \theta_0) \in \mathbb{Z}.
\end{equation}
The gauge transformation $\theta \mapsto \theta + 2\pi$ is accompanied by a monodromy permuting the branches, $j \mapsto j + n$.\footnote{This is an instance of the standard result that Chern-Simons terms dualize to Stueckelberg terms, in a slightly non-standard form where $j$ is viewed as a field strength even though there is no obvious way to define a $(-1)$-form gauge field for it to be the field strength of. Readers can find a pedagogical review of Chern-Simons terms, Stueckelberg terms, and dualities between them in~\cite{Reece:2023czb}.} This theory can admit domain walls, electrically charged under $A_3$, across which the value of $j$ jumps. We refer to these domain walls as $A$-branes. A non-canonical kinetic term for $A_3$ would lead to a non-quadratic effective potential for $\theta$; while this may seem to go beyond the validity of EFT, these terms may be determined by some organizing principle like the DBI action that nonlinearly realizes higher-dimensional Poincar\'e invariance~\cite{Silverstein:2008sg}. Useful earlier references related to four-form gauge field strengths, associated domain walls, and effects on axion mass include~\cite{Brown:1987dd, Brown:1988kg, Gabadadze:1999na, Bousso:2000xa, Dvali:2005an}.

\subsection{Making the axion monodromy mass dynamical}
\label{subsec:dynamicalinteger}

An axion monodromy mass can be very large; for instance, both $f$ and $e_A$ are plausibly of order a fundamental UV cutoff like the string scale. Such a mass could easily suppress axion fluctuations during inflation. The question, then, is how to turn it off to allow the axion to solve the Strong CP problem and acquire a dark matter abundance at late times. For this, we need $n$ to be a dynamical integer.

A dynamical integer is one that can change discontinuously across a dynamical domain wall, like the integer $j$ in~\eqref{eq:jF4relation}. We can interpret $n$ as dynamical by viewing it as the electric flux of a {\em different} 3-form gauge field $B_3$: $n \sim -\frac{1}{e_B^2}{\star \rmd B_3}$. More precisely, because the field strength $n$ participates in the Chern-Simons interaction~\eqref{eq:thetaF4}, when we dualize to a 3-form gauge theory we find a kinetic term of Stueckelberg type,
\begin{equation} \label{eq:B3kinetic}
-\int \frac{1}{2e_B^2} \left[\rmd B_3 - \frac{1}{2\pi}(\theta - \theta_0) F_4\right] \wedge \star \left[\rmd B_3 - \frac{1}{2\pi}(\theta - \theta_0) F_4\right].
\end{equation}
In this duality frame, the generator of the $\mathbb{Z}$ gauge symmetry associated with the axion periodicity also acts on the 3-form gauge field:
\begin{equation}\label{eq:thetaA3transform}
    \theta \mapsto \theta + 2\pi, \quad B_3 \mapsto B_3 + A_3.
\end{equation}
We denote the gauge invariant field strength as
\begin{equation}
    \widetilde{H}_4 \equiv \rmd B_3 - \frac{1}{2\pi}(\theta - \theta_0) F_4.
\end{equation}
The precise relationship between $B_3$ and $n$ is then
\begin{equation} \label{eq:nH4relation}
   n = -\frac{1}{e_B^2} {\star\widetilde{H}_4} \in \mathbb{Z}.
\end{equation}

\begin{figure}[htbp]
    \centering
    \includegraphics[width=0.6\textwidth]{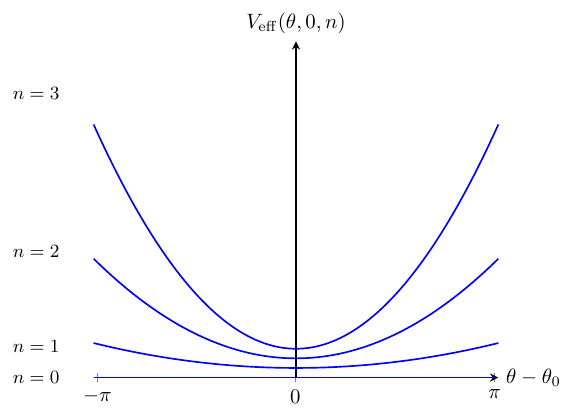}
    \caption{The axion monodromy potential for different flux levels $n$, for fixed $j=0$. The potential has the form $V_{\mathrm{eff}}(\theta,0,n) = \frac{1}{2} n^2 \left[\frac{e_A}{2\pi}(\theta - \theta_0)\right]^2 + V_n$. Our solution to the axion isocurvature problem relies on a cosmological phase transition connecting the $n = 1$ branch with the $n=0$ branch, effectuated by dynamical $B$-branes. The constant offset $V_n$ can differ for different choices of $n$.}
    \label{fig:parabola-vary-N}
\end{figure}

For fixed $n$, this modified theory gives rise to essentially the same effective potential~\eqref{eq:Axion-Effective-Potential} discussed above, except that there can be a $\theta$-independent but $n$-dependent constant term $V_n$:
\begin{equation}
V_\mathrm{eff}(\theta, j, n) = \frac{1}{2} n^2 \left(\frac{e_A}{2\pi}\right)^2  \left(\theta - \theta_0 - \frac{2\pi j}{n}\right)^2 + V_n.
\label{eq:Axion-Effective-Potential-2}
\end{equation}
The $n$-dependence of the effective potential is illustrated in Fig.~\ref{fig:parabola-vary-N}.
Minimally, we might expect $V_n \propto n^2$ from the kinetic term~\eqref{eq:B3kinetic}, but the $n$-dependence could take a more complicated form in the presence of higher powers of $\widetilde{H}_4$ in the effective action.

Because $n$ is an electric flux, its value changes across objects that carry electric charge under $B_3$, that is, domain walls or 2-branes. We will refer to these objects as $B$-branes. The effective action on the worldvolume $M$ of a unit-charge $B$-brane takes the form
\begin{equation}
    S_B = \int_M \sqrt{|g_B|}\, \rmd^3x\,{\cal T}_B + \int_M B_3 + \int_M \frac{\theta - \theta_0}{2\pi}(\rmd b_2 - A_3).
\end{equation}
The first term is the brane tension: the $B$-brane has a mass per unit area equal to ${\cal T}_B$. The second term is the electric coupling of the $B$-brane to $B_3$. The third term is needed to compensate the nontrivial gauge transformation of $B_3$ from~\eqref{eq:thetaA3transform}. This is a form of anomaly inflow~\cite{Callan:1984sa}: $b_2$ is a brane-localized 2-form gauge field that shifts under an $A_3$ gauge transformation. This allows $B$-branes to carry dissolved $A$-brane charge.

The $A$-brane can end on an axion string. In a region with $n = 0$, the axion strings are free; in the region with $n = 1$, they are confined, with a segment of $A$-brane stretching out of the $B$-brane when we push a string through.

The $B$-brane is the central dynamical player in our mechanism. A cosmological phase transition between the $n \neq 0$ early universe and the $n = 0$ late universe proceeds by the nucleation of $B$-brane bubbles, which must expand and merge for the phase transition to complete. We will discuss this process further in \S\ref{sec:cosmology}.

\subsection{UV cutoff and constraints on the Hubble scale}

In the following sections, we will work with a four dimensional EFT of the axion and inflaton which has an associated UV cutoff. Since we are assuming an extra-dimensional axion (as we will discuss in more detail in \S\ref{sec:models}), the most obvious cutoff to consider is the Kaluza-Klein (KK) scale $M_{\rm KK}$, whose value depends on the type of UV completion. The other possible cutoff is the quantum gravity cutoff (or species scale) $\Lambda_\mathrm{QG}$, which in string theory setting is essentially the string scale $M_s$. Both scales can vary depending on the precise UV completion, but given that our EFT is inspired by string theory models, we will look to those in order to judge both the KK scale and the string tension. The bound $H_I < M_\mathrm{KK}$ is always stronger than the bound $H_I < \Lambda_\mathrm{QG}$, but the latter bound can be quantified in terms of axion parameters in a less model-dependent manner.

The tension $\mathcal{T}$ of an axion string provides an upper bound on the quantum gravity scale, $\Lambda_\mathrm{QG}^2 \lesssim 2\pi \mathcal{T}$, if the axion string originates from a fundamental object (e.g., a D-brane or F-string) wrapped on a cycle with volume large in string units. As argued in~\cite{Reece:2025thc, Reece:2025zva}, there is often an axion string tension that parametrically exhibits the electric-magnetic ``co-scaling'' relationship 
\begin{equation}
    \mathcal{T}\approx 2\pi S_{\rm inst} f^2,  
\end{equation}
where $S_{\rm inst}=8\pi^2/g_{\rm YM}^2$ is the instanton action of the QCD sector. This leads to the bound
\begin{equation} \label{eq:LambdaQGbound}
   \Lambda_\mathrm{QG} \lesssim 2\pi \sqrt{S_\mathrm{inst}} f.
\end{equation}
Although exceptions to co-scaling are known, for instance near conifold points in moduli space~\cite{Svrcek:2006yi, Benabou:2025kgx, Reece:2025zva}, other arguments based on naturalness or unitarity~\cite{Reece:2025thc, Seo:2024zzs, Benabou:2025kgx} provide independent arguments for~\eqref{eq:LambdaQGbound}, and we are not aware of exceptions to it for fundamental, extra-dimensional axions. As a rough estimate we can approximate $2\pi \sqrt{S_{\rm inst}}\approx 80$ where, as is conventional, we have assumed the value of the gauge coupling at the GUT scale $\alpha_{\rm YM}\equiv g_{\rm YM}^2/(4\pi) \approx 1/25$.

Requiring that the Hubble scale is below the quantum gravity cutoff then imposes the constraint
\begin{equation}
    \label{eq:string_tension_bound}
    H_I \ll \Lambda_\mathrm{QG} \lesssim 80 f,
\end{equation}
which is weaker than the bound for the standard (4d) Peccei-Quinn axion.\footnote{The PQ symmetry breaking scale is order $f$, and requiring that it is broken during inflation implies $H_I \ll f$, up to order one factors. To restore the symmetry, fluctuations of the axion must be sufficient to have it wind around its field space, whose scale is set by $f$, which produces a string. Since during inflation the massless axion has Hubble sized fluctuations the bound is $H_I \ll f$. This is consistent with the string tension scale since a PQ axion string has tension $\mathcal{T}\approx \pi f^2$.} Generally in string theory models the dependence of the KK scale with the volume of internal dimensions is
\begin{equation}
    M_{\rm KK} \sim \frac{M_s}{\mathcal{V}^{1/6}}
    \label{eq:M_KK}
\end{equation}
where $\mathcal{V}$ is the volume in units of the string length $\ell_s$ and the string scale $M_s\equiv 2\pi/\ell_s$. In order to compare the two we note that the string theory is perturbative only when $\mathcal{V} \gg 1$ and $g_s \lesssim 1$ which requires
\begin{equation}
    M_s^2 \approx \frac{g_s^2}{\mathcal{V}} M_{\mathrm{Pl}}^2 \ll M_{\mathrm{Pl}}^2.
    \label{eq:M_s}
\end{equation}
It is also necessary to ensure $g_s \sim 1$ because the gauge coupling $g_{\rm YM} \sim g_s$ is observed to be order one. Upon requiring the gauge coupling to be order one, the string scale is typically of order the upper bound~\eqref{eq:LambdaQGbound}, so that $f \sim 10^{-2} M_s \sim 10^{-2} M_\mathrm{Pl}/\sqrt{\mathcal{V}}$ and we obtain the hierarchy
\begin{equation}
    f \lesssim M_{\rm KK} \lesssim M_s \ll M_{\mathrm{Pl}}\,,
\end{equation}
which can be met by having $M_s \sim 10^{14} \, {\rm GeV}$, $f \sim 10^{12}\,{\rm GeV}$ and $M_{\rm KK}\sim 10^{13}\,{\rm GeV}$, which can be engineered in e.g.~a Type IIB theory compactified on a 6d Calabi-Yau orientifold \cite{Conlon:2006tq}.\footnote{In this case, we have the freedom to fix $M_s \sim 10^{14}\,{\rm GeV}$ because because the QCD gauge coupling also depends on the volume of the 4-cycle wrapped by a D7 brane. We do not have this luxury in e.g.~heterotic string theory where the gauge coupling fixes $M_s \approx \frac{\sqrt{\pi}}{5} M_{\mathrm{Pl}}$ and, requiring $g_s \approx 1$, $\mathcal{V}\approx 25$.} In this scenario the KK scale sets the cutoff and we require $H_I \ll 10^{13}\, {\rm GeV}$, which is more stringent than the bound from BICEP-Keck. 

%%%%%%%%%%%%%%%%%%%%%%%%%%%%%%%%%%%%%%%%%%%%		
\section{Cosmological History}
\label{sec:cosmology}

The essential idea of our mechanism is that the integer $n$ in~\eqref{eq:thetaF4} is nonzero in the early universe (say $n = 1$, for simplicity) and zero in the late universe, with a first-order phase transition in between. There are a number of different ways of embedding this mechanism into the cosmological history of the universe. The phase transition must occur after the period of inflation that gives rise to the perturbations observed in the CMB, since axion isocurvature perturbations are absent at those scales. We assume that it occurs before the axion begins to oscillate (around the time of the QCD phase transition), so as not to disturb the production of axion dark matter. This leaves a wide range of possibilities.

\subsection{Bubble nucleation and percolation}

Regardless of when the phase transition happens, it must proceed by the nucleation of $B$-brane bubbles. Across a $B$-brane, the flux $n$ changes by one unit. Unlike the standard Coleman scenario of tunneling by scalar fields~\cite{Coleman:1977py,Callan:1977pt,Coleman:1980aw}, the bubble wall is not simply a configuration made up of a scalar field gradient, it is an object that carries a charge (under the gauge field $B_3$). This is the scenario of {\em flux tunneling}, as discussed in, e.g.,~\cite{Blanco-Pillado:2009lan,Brown:1987dd,Bousso:2000xa,Brown:2010bc,Brown:2010mg,Brown:2016nqt}.\footnote{Here we are referring to the nature of the EFT in which the bubble walls are charged under a 3-form gauge field, but we will see in Sec.~\ref{sec:models} that the UV completion may involve an extra-dimensional flux as well.} 

As in the standard formalism, the vacuum decay rate per unit volume and unit time $\Gamma$ is estimated as $\Gamma \sim m_B^4 \E^{-S_E}$, where $S_E$ is the Euclidean bounce action and $m_B^4$ a functional determinant related to the positive perturbation modes living on the brane, parametrized by a typical energy scale $m_B$. The vacuum persistence rate $p(t)$, which measures the probability that a point in space remains in the false vacuum at physical time $t$, is given by~\cite{Guth:1980zm,Guth:1982pn} 
\begin{equation}
    p(t) \approx \E^{-\frac{4\pi}{3}\gamma H_I(t-t_i) }.
\end{equation}
Here $\gamma = \frac{\Gamma}{H_I^4}$ is called the nucleation efficiency, which measures the number of bubbles generated per Hubble time and Hubble volume.

Inside one Hubble patch, quantum tunneling creates bubbles of the $n=0$ vacuum at the critical radius $\bar{\rho}$. We can treat the $B$-brane as an infinitely thin object of fixed tension, ignoring its couplings to scalar fields, for a ``thin brane'' version of the standard thin-wall estimate of Coleman and De Luccia~\cite{Coleman:1980aw}. For tunneling from the $n =1$ to $n=0$ vacuum, the critical bubble radius $\overline{\rho}$ and instanton action $S_E$ are given by
\begin{equation} 
    \bar{\rho} = \frac{\rho_0}{1+\left(\frac{\rho_0 H_I }{2}\right)^2}, \quad \quad S_E=\frac{S_0}{1+\left(\frac{\rho_0 H_I }{2}\right)^2},
    \label{eq:critical-bubble}
\end{equation}
with
\begin{equation} \label{eq:thinwallest}
       \rho_0 = \frac{3\sigma}{\Delta V} \quad \quad S_0 = \frac{27\pi^2}{2} \frac{\sigma^4}{(\Delta V)^3}.
\end{equation} 
Here $\sigma$ is the brane tension and $\Delta V$ is the vacuum energy difference between the $n = 1$ and $n = 0$ vacua.

%%%%%%%%%%%%%%%%%%%%%%%%%%%%%%%%%%%%%%%%%%%%%%%%%%%%%%%

\begin{figure*}[!t]\begin{center}
\includegraphics[width=0.95\textwidth]{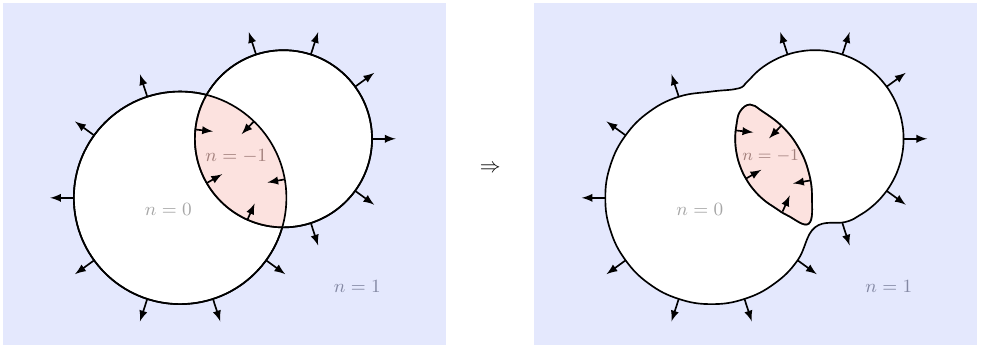} 
\end{center}
\caption{
The process of $B$-brane bubble mergers. As explained in~\cite{Blanco-Pillado:2009lan}, when bubbles merge, the overlap region will have a different flux ($n = -1$ in this example). Left panel: the state immediately after bubbles collide. Arrows indicate the direction of pressure on the bubble walls due to vacuum energy differences, assuming that the $n = 0$ state has lowest vacuum energy. Right panel: we expect that the branes will reconnect and interior regions with $n \neq 0$ will collapse, eventually leading to a universe in the $n = 0$ state everywhere.
}
\label{fig:merger}
\end{figure*}%
%%%%%%%%%%%%%%%%%%%%%%%%%%%%%%%%%%%%%%%%%%%%%%%%%%%%%%

The critical radius $\bar{\rho}$ is always smaller than the comoving Hubble radius $H_I^{-1}$. If only a single bubble nucleates, the expansion rate of the bubble wall will soon accelerate towards the speed of light. To an outside observer, the bubble wall will eventually move along a null geodesic and enclose a causal Hubble patch. Vacuum energy fully transforms into the kinetic energy of bubble wall and fails to ignite a reheating process~\cite{Guth:1982pn}. If more than one bubble nucleates inside the same Hubble patch, these bubbles will expand and merge into a single true vacuum bubble. 

An important physical effect in flux tunneling, emphasized in~\cite{Blanco-Pillado:2009lan}, is that because the value of $n$ changes across a $B$-brane, the merger of bubbles creates an intermediate state with a different value of $n$, as illustrated in the left panel of Fig.~\ref{fig:merger}. If the phase transition is between an $n = 1$ state and an $n = 0$ state, the merger of two bubbles produces an intermediate bubble of the $n = -1$ state. We assume, crucially, that the $n = 0$ state (in which we live today) has lower energy than all $n \neq 0$ states.\footnote{Much of the literature on this topic assumes Freund-Rubin-like compactifications, in which the flux being studied is balanced against curvature to stabilize the extra dimensions. Such cases are incompatible with our assumption of a minimum (and finite) energy $n = 0$ state. We have in mind a UV completion where the flux $n$ is through a small cycle within a larger compactification and has little effect on the overall volume, so that we do not have to make the same assumptions about the $n$-dependence of $V_n$ made in other works.} It is then energetically favored for the bubble walls to reconnect, allowing the $n = -1$ region to pinch off and collapse, as in the right panel of Fig.~\ref{fig:merger}. As it collapses, it will radiate away energy into light fields. If the phase transition occurs at or after the end of inflation, the collapse of such negative-$n$ bubbles could play an important role in reheating, though it is difficult to make a more precise statement without a specific UV completion. The collision of bubble walls could also produce primordial gravitational waves, which may provide a detectable signal of a first-order phase transition. We will briefly comment on this in our concluding section~\ref{para:GR-signal}.

\subsection{4d Effective field theory set-up}

In \S\ref{sec:mechanism}, we explained the ingredients of an effective field theory in which the axion mass is controlled by a dynamical integer. In the cosmological context, another important ingredient is the inflaton field $\phi$. Thus, we should consider a more general EFT, in which the inflaton and axion are coupled. There are multiple reasons for this. One, which we will discuss in more detail below, is that a consistent cosmology in which tunneling occurs during inflation requires that the tunneling rate be time-dependent, which is most easily accomplished if the parameters controlling tunneling are $\phi$-dependent. A more basic reason is simply that EFT allows such couplings, so there is no reason not to write them.

The 4d effective field theory in which we can calculate a tunneling rate using the thin-brane approximation has as dynamical quantities the inflaton field $\phi$, the axion field $\theta$, the dynamical integer $j$ (which we fix to $j = 0$ as it plays no role in our cosmology), the dynamical flux integer $n$, and the worldvolume $M$ of the $B$-brane across which the flux drops from $n + 1$ to $n$. We parametrize the action in general as
\begin{align} \label{eq:inflatonEFT}
S = & \int \sqrt{|g|}\mathrm{d}^4 x \left[ -\frac{Z_n}{2}(\partial \phi)^2 -\frac{1}{2}f_n(\phi)^2 (\partial \theta)^2 - V(\phi, \theta, n)\right] \nonumber - \int_M {\cal T}_n(\phi)\sqrt{|g|}\mathrm{d}^3 x, \nonumber \\
V(\phi, \theta, n) = & V_n(\phi) + \frac{1}{2} n^2 \left(\frac{e_A(\phi)}{2\pi}\right)^2  \left(\theta - \theta_0\right)^2.
\end{align}		
The $\phi$ dependence of various coefficients in the action is expected for modulus fields that describe the geometry of extra dimensions.

Before the phase transition, $n \neq 0$; for simplicity we assume $n = 1$. We assume that the axion mass $\frac{n}{2\pi f} e_A(\phi) > \frac{3}{2}H_I$ for the full range of $\phi$ relevant for inflation, so that the axion field remains relaxed to the minimum of its potential with highly suppressed perturbations (recall Eq.~\eqref{eq:massivemodefunction}). We assume that the inflaton potential $V_1(\phi)$ supports slow roll inflation. Within this effective field theory, the simplest estimate that one can make of the tunneling rate is the thin-wall, thin-brane approximation (Eqs.~\eqref{eq:critical-bubble}, \eqref{eq:thinwallest}) using the instantaneous values of $\sigma$ and $\Delta V$ at a given time, that is, $\sigma(t) = {\cal T}_0(\phi(t))$, $\Delta V(t) = V_1(\phi(t)) - V_0(\phi(t))$. 

For a problem of flux tunneling between two extrema of the potential, one could improve upon this simple instantaneous thin-brane approximation by solving for a bounce solution for the scalar field $\phi$ and the metric in the presence of the $B$-brane bubble wall, with appropriate boundary conditions to match the solution inside and outside the brane. It would be interesting to pursue the detailed derivation of such boundary conditions, and a more complete tunneling calculation, in future work. The problem of tunneling during inflation has the additional complication that the initial state is not a vacuum, but a slowly rolling solution. Refining the tunneling formalism to such a case is an interesting theoretical problem. On the other hand, for the interesting case of nucleation efficiency $\gamma \gg 1$ so that the phase transition completes, it should be self-consistent to neglect the slow time dependence of the initial state.

\subsection{Dynamical nucleation efficiency}

If the tunneling process occurs during inflation, the nucleation efficiency $\gamma$ cannot be constant throughout the inflation era, otherwise we would encounter ``the graceful exit'' problem of old inflation~\cite{Guth:1980zm,Guth:1982pn,Hawking:1982ga}. When $\gamma \gg 1$, the phase transition proceeds very rapidly. The timescale for bubbles to nucleate and percolate is small, ending inflation early compared to the duration of inflation that is required to resolve cosmological problems.  When $\gamma \ll 1$, on the other hand, the percolation rate is negligible, the phase transition proceeds slowly, and the universe never reheats. While this has typically been resolved with the new inflation paradigm, an alternative elegant way to circumvent the graceful exit problem is making $\gamma$ dynamical, by coupling an inflaton field with another spectator field responsible for tunneling. This is the paradigm of double field inflation~\cite{Adams:1990ds,La:1989za,Linde:1990gz}. 

In our case, we instead make $\gamma$ dynamical by coupling the inflaton field with the magnetic brane through the interaction term in~\eqref{eq:inflatonEFT},
\begin{equation}
    S \supset \int_{M} \mathcal{T}_n(\phi) \sqrt{|g|}\rmd^3x, 
\end{equation}
where $M$ is the worldvolume of magnetic brane. During inflation, the inflaton slowly rolls, $\mathrm{d}\phi/\mathrm{d}t \neq 0$, and the instanton action $S_E(t)$ becomes time dependent. We can choose the tension $\mathcal{T}_n(\phi)$ to be monotonically decreasing, so that $\gamma$ could be small in early inflation and become sufficiently large later. The condition for a first order phase transition to complete at some physical time $t_0$ is given by (following Appendix A of~\cite{An:2020fff})
\begin{equation}
    S_E(t_0) \simeq \log {\frac{m_B^4}{\beta^4(t_0)}},
\end{equation}
where $\beta(t) \equiv -\rmd S_E / \rmd t$ is the first time derivative of the time-dependent instanton action. The above expression is only valid when $S_E(t_0) > 0$, and hence $m_B > \beta(t_0)$. In order for the tunneling process to complete, we require that $\gamma(t_0) \gg 1$ and also $\frac{\beta(t_0)}{H_I(t_0)} \gg 1.$ 

We assume that the dominant time dependence is in $\mathcal{T}_n(t)$, and use ${\dot{\mathcal{T}}}_n(t) = \frac{\rmd \mathcal{T}_n}{\rmd \phi(t)} {\dot \phi}(t)$, which we can express in terms of the first slow-roll parameter $\epsilon(t)$ using ${\dot \phi}(t) = \sqrt{2 \epsilon(t)} H_I(t) M_\mathrm{Pl}$. We further define the dimensionless quantity $\varepsilon_\rho \equiv \rho_0(t_0) H_I(t_0) = 3 \sigma(t_0) H_I(t_0)/\Delta V(t_0)$, which appears in the Coleman-De Luccia estimate~\eqref{eq:thinwallest}, and which we assume to be small. 
Expressed in our model parameters, we have
\begin{align}
    \frac{\beta(t_0)}{H_I(t_0)} & \approx \frac{2\pi^2}{H_I(t_0)^3} \sqrt{2\epsilon(t_0)}M_\mathrm{Pl} \varepsilon_\rho^3 \frac{1+\frac{\varepsilon_\rho^2}{8}}{\left(1+\frac{\varepsilon_\rho^2}{4}\right)^2} \left|\frac{\rmd \mathcal{T}_n}{\rmd \phi}(t_0)\right| \nonumber \\
    & \approx 7 \times 10^4 \frac{1}{H_I(t_0)^2} \varepsilon_\rho^3 \left|\frac{\rmd \mathcal{T}_n}{\rmd \phi}(t_0)\right| \gg 1.
    %\frac{54\pi^2}{1+\alpha}\frac{{\mathcal{T}_n}^3(t_0)\frac{\rmd {\mathcal{T}_n}}{\rmd \phi}(t_0)}{(\Delta V)^3} M_\mathrm{Pl}\sqrt{2\epsilon(t_0)} \gg 1.
\end{align}
In the last step, we have used $\sqrt{2 \epsilon(t)} M_\mathrm{Pl} = \frac{H_I(t)}{2\pi\sqrt{A_s(t)}}$ with the measured Planck value of the Planck spectral index (see~\eqref{eq:scalarpower}) to obtain the numerical prefactor $\pi/\sqrt{A_s} \approx 7 \times 10^4$. This estimate shows that  ${\rmd \mathcal{T}_n}/{\rmd \phi}$ cannot be too small, in Hubble units.

Depending on the timing of the phase transition relative to the end of inflation, three possible cosmological histories can be considered.
\begin{itemize}
    \item {\bf Tunneling during inflation.} The vacuum energy drops partially during the phase transition, but still drives a second inflation stage with a lower expansion rate $H_2 < H_1$. The monodromy mass of the axion turns off during the phase transition, so isocurvature perturbations are generated in the second epoch of inflation. These non-adiabatic perturbations from the later stage of inflation will re-enter the horizon after reheating much earlier than the CMB scale perturbations observed today, so they give rise to isocurvature perturbations on small scales. In this scenario, both the inflaton field and the axion field experience a ``kick'' during the phase transition, which gives rise to interesting features in both the adiabatic and isocurvature power spectrum. We will study this case in detail in the next section.

    \item {\bf Tunneling to end inflation.} The phase transition eliminates both the axion monodromy mass and the inflationary vacuum energy, after which the universe starts reheating and thermal expansion. The nucleation efficiency $\gamma$ must be small enough to support approximately 60 e-folds of inflation, while sufficiently large later to gracefully exit inflation. The monodromy potential keeps the axion field heavy throughout inflation, so the axion isocurvature constraints from Eqs.~\eqref{eq:isocurvconstraint},~\eqref{eq:axionisocurv} are removed.

    \item {\bf Tunneling after inflation.} Another possibility is that the tunneling rate $\gamma$ remains small throughout inflation, so that the phase transition completes only after inflation ends, when the inflaton rapidly rolls down the slope of its potential, lowering $H$. Because the initial state in this case is further from approximate de Sitter space, it may require a larger departure from the standard Coleman tunneling formalism. We will not consider this case in detail, but note that it has some similarities with a recent discussion in~\cite{Buckley:2024nen}. As with the case of tunneling to end inflation, the axion isocurvature constraint is eliminated in this scenario.
\end{itemize}

%%%%%%%%%%%%%%%%%%%%%%%%%%%%%%%%%%%%%%%%%%%%		
\section{Case Study: Tunneling During Inflation}
\label{sec:casestudy}

In this section we will determine the evolution of the axion and the inflaton through two phases of inflation. The first phase of inflation is characterized by a Hubble scale $H_1$, during which the axion has a mass $m_\theta > \frac{3}{2}H_1$. In the second phase the Hubble constant becomes $H_2<H_1$ and the axion becomes massless. We will often also use the dimensionless parameter $\nu_\theta \equiv \sqrt{(m_\theta/H_1)^2 - 9/4}$ instead of the mass. The metric for this period is written as
\begin{equation}
    \rmd s^2 = -\rmd t^2 + a^2(t) \rmd \mb{x}^2
\end{equation}
During the first phase of inflation, the scale factor is $a(t)=\E^{H_1 t}$ and it continuously transitions into the second phase at $t_c$ where $a(t)=\E^{H_2(t-t_c)+H_1 t_c}$. For our purposes, it will be sufficient to work in the de Sitter limit throughout. We will generally ignore slow-roll corrections while determining the evolution of the mode functions of both the inflaton and the axion, and only keep track of it through normalization of the co-moving curvature perturbation. 

We will find it convenient to work with the conformal time variable $\conft = -H_1^{-1}\E^{-H_1 t}$ so that the phase transition occurs at $\conft_c = -H_1^{-1}\E^{-H_1 t_c}$. This is the conformal time of the first phase, but \textit{not} the second phase. In this time coordinate the scale factor is
\begin{equation}
    a(\conft) = \begin{cases}
        -\frac{1}{H_1 \conft} \,,&\conft<\conft_c \\
        -\frac{1}{H_1 \conft_c}\left(\frac{\conft_c}{\conft}\right)^{\hratio} \,,&\conft>\conft_c
    \end{cases}
\end{equation}
where $\hratio\equiv H_2/H_1 <1$ measures the change in the Hubble scale through the phase transition.

Our main goal in this section is to determine the inflationary axion power spectrum, and subsequently the axion dark matter isocurvature.

Let us first establish some generalities. We will be dealing with a scalar operator $\hat{X}_{\mb{k}}(\conft)$ with co-moving momentum $\mb{k}$, where $\hat{X} \in \{\delta\hat{\theta}, \delta\hat{\phi}\}$ is either the axion or the inflaton fluctuations, and assume that the state for each is its standard Bunch-Davies vacuum. The operator can be written in terms of its mode functions
\begin{equation}
    \hat{X}_{\mb{k}}(\conft) = X_{k}(\conft) \hat{a}_{\mb{k}} + X_{k}^{*}(\conft) \hat{a}_{-\mb{k}}^{\dagger}
\end{equation}
and the vacuum $|\Omega\rangle$ is defined by the condition $\hat{a}_{\mb{k}} | \Omega \rangle = 0$. The equal time power spectrum for an operator $\hat{X}_{\mb{k}}(\conft)$ is defined as
\begin{equation}
    \langle \hat{X}_{\mb{k}}(\conft) \hat{X}_{\mb{k}}^*(\conft)\rangle = (2\pi)^3\delta_{\rm D}(\mb{k}+\mb{k}') P_{X}(\conft, k)
\end{equation}
and the power spectrum is the norm squared of the mode functions
\begin{equation}
    P_X(\conft,k) = |X_k(\conft)|^2\,.
\end{equation}
We will first compute the axion power spectrum in this background and discuss the inflationary scalar power spectrum in appendix~\ref{app:inflaton_fluct}. Let us begin with the axion.

\subsection{Axion power spectrum and isocurvature}\label{sec:axion_fluct}
The axion mode function $\delta\theta_{k}(\conft)$ first goes through a period of evolution determined by its mass $m_\theta$. Our goal will be to determine the evolution of the axion mode function, assuming that it begins in the standard Bunch-Davies vacuum. Before embarking on the computation, let us qualitatively understand this scenario. In general, a mode with momentum $k$ will learn that it is massive when $k/a \approx m_\theta$ or $\conft \approx -\nu_\theta/k$, where we have assumed $m_\theta \gg H_I$. This crossing of the `mass-horizon' occurs earlier for long-wavelength modes and later for short-wavelength ones. Therefore, we expect that by the end of inflation, the small $k$ power spectrum will essentially be that of a massive field, and therefore sensitive to the mass. At large $k$, however, we expect that the mode function never `learns' that it is massive and is therefore unaffected by the phase transition. So we expect that on short scales, we will recover the standard scale-invariant power spectrum exhibited by a massless field. 

At which scales should we expect to see the scale-invariant spectrum? We can estimate this scale as follows: a mode with momentum $k$ will not learn that it is massive until $\conft \approx -\nu_\theta/ k$. If $\conft>\conft_c$, we should expect this mode to remain effectively massless throughout inflation. We refer to the scale which was horizon sized at the time of the phase transition as the critical scale $k_c$, i.e.,
\begin{equation} \label{eq:defkc}
k_c \conft_c = -1.
\end{equation}
Then in terms of momenta, the effectively massless modes at the time of the phase transition have $k > \nu_\theta k_c$. This tells us that we should expect the scale-invariant power spectrum to kick in after 
\begin{equation} \label{eq:defkt}
    k_t \approx \nu_\theta k_c,
\end{equation}
i.e., for scales which are smaller than the critical scale.

With the generalities at hand, let us now derive the precise axion power spectrum. During the initial massive phase the mode function $\delta\theta_k(\conft)$ obeys the standard wave equation in de Sitter
\begin{equation}
    \delta\theta_k''(\conft) -\frac{2}{\conft}\delta\theta_k'(\conft) + \left(k^2 + \frac{m_\theta^2}{H_1^2 \conft^2}\right)\delta\theta_k(\conft) = 0,
\end{equation}
where the prime denotes a derivative with respect to conformal time $\conft$. 
In the Bunch-Davies vacuum this admits the solution (recalling~\eqref{eq:massivemodefunction})
\begin{equation}
    \delta\theta_k(\conft) = \frac{\sqrt{\pi}}{2 f_a} H_1 \E^{-\frac{1}{2}\pi \nu_\theta}\E^{i \frac{\pi}{4}}(-\conft)^{\frac{3}{2}} \mathrm{H}_{i \nu_\theta}^{(1)}(-k \conft)\,.
\end{equation}
This will set the initial conditions for the evolution of the axion during the massless phase. For this second phase the equation of motion is
\begin{equation}
    \delta\theta_k''(\conft) + \frac{(1-3\hratio)}{\conft}\delta\theta_k'(\conft) + k^2 \left(\frac{\conft_c}{\conft}\right)^{2-2\hratio} \delta\theta_k(\conft) = 0,
\end{equation}
which takes this unfamiliar form because we have chosen to work with the conformal time corresponding to the first phase of inflation. This equation admits the following general solution:
\begin{equation}
    \begin{aligned}
        \delta\theta_k(\conft) &= A_k \left[1+ i k \conft_c \hratio^{-1}\left(\frac{\conft}{\conft_c}\right)^{\hratio}\right] \exp\left[-i k \conft_c \hratio^{-1}\left(\frac{\conft}{\conft_c}\right)^{\hratio}\right] \\
        &+B_k \left[1- i k \conft_c \hratio^{-1}\left(\frac{\conft}{\conft_c}\right)^{\hratio}\right] \exp\left[i k\conft_c \hratio^{-1}\left(\frac{\conft}{\conft_c}\right)^{\hratio}\right],
    \end{aligned}
\end{equation}
where $A_k$ and $B_k$ are arbitrary coefficients to be fixed by initial conditions, i.e., by matching across the phase transition.\footnote{The mode function looks slightly unusual compared to the standard massless mode functions. This is entirely because of our definition of conformal time $\conft$, which is chosen to match the standard conformal time of the first inflationary phase.} The matching conditions, i.e.~continuity of $\delta\theta_k(\conft)$ and $\delta\theta_k'(\conft)$ across the phase transition at $\conft_c$, can be determined through the equation of motion for the axion. Usually, the axion is chosen to be in its ground state which is synonymous with setting $B_k = 0$. We will see that the sudden turning off of the mass will instead displace the axion away from its Bunch-Davies state and produce a $B_k \neq 0$. This is typically expected in cosmological scenarios with phase transitions (see e.g.~\cite{DAmico:2020euu}).

\begin{figure}
    \centering
    \includegraphics[width=0.95\textwidth]{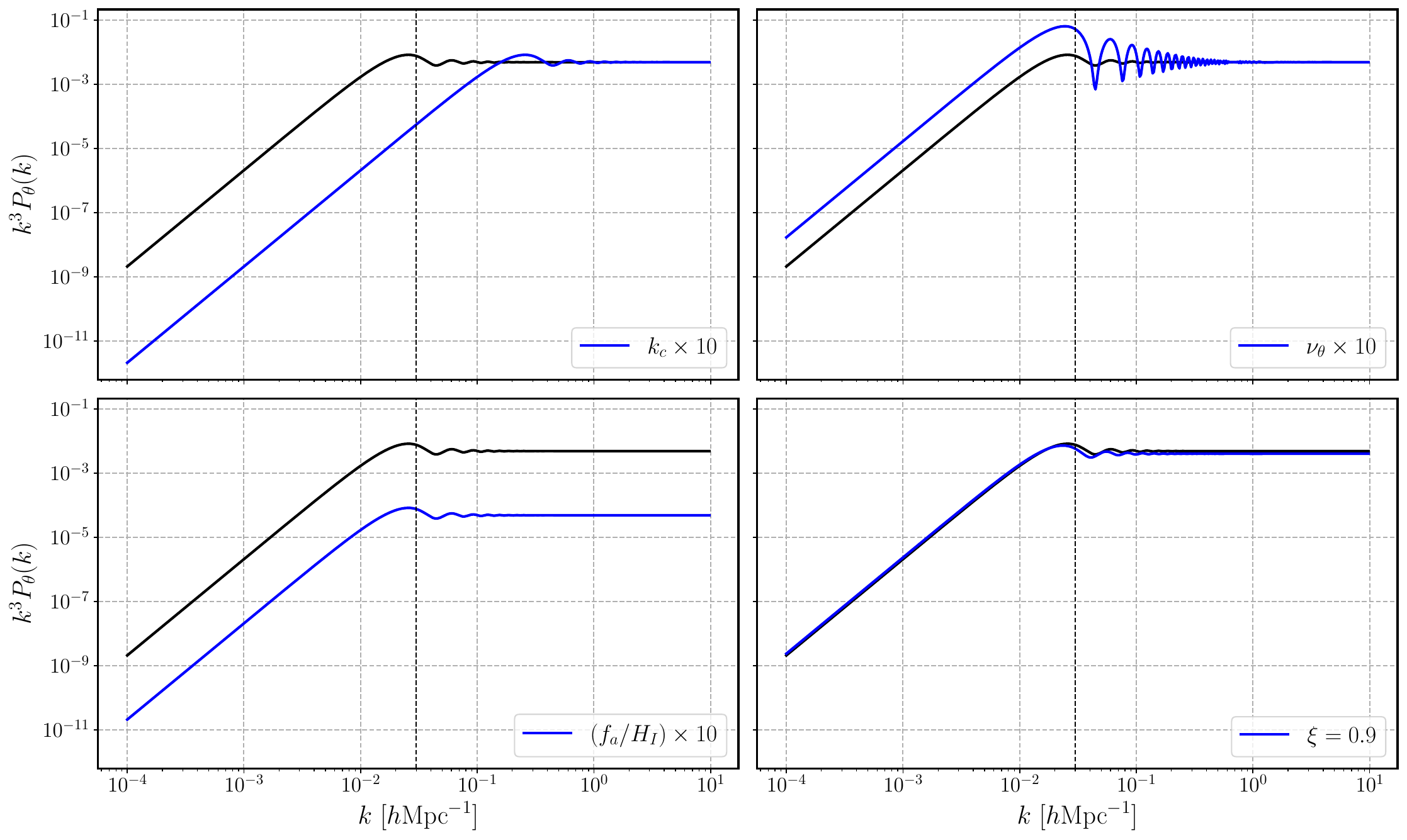}
    \caption{We plot the axion power spectrum $P_\theta(k)$ and show how its response as we change each parameter that determines its shape. For reference, we mark $k_c$ for the base case in the black dashed line. The base set of parameters are $k_c=10^{-2}\,h {\rm Mpc}^{-1}$, $\nu_\theta=3$, $(f_a/H_I)=10$ and $\hratio=0.99$, which is shown in solid black in each subplot. The variation in each parameter is shown sequentially in blue.}
    \label{fig:axion_pk}
\end{figure}

Using the matching conditions we can determine the unknown coefficients to be
\begin{equation}
    \begin{aligned}
        A_k &= \frac{i\pi^{\frac{1}{2}}\hratio^{2}}{4 f_a k^2}H_1 \E^{-\frac{\pi \nu_\theta}{2}}\E^{i \frac{\pi}{4}}(-\conft_c)^{-\frac{1}{2}}\E^{i k \conft_c \hratio^{-1}}(1- i k \conft_c \hratio^{-1}) \mathrm{H}_{1+ i \nu_\theta}^{(1)}(-k \conft_c) \\
        &-\frac{\pi^{\frac{1}{2}}\hratio^{2}}{8 f_a k^3}H_1 \E^{-\frac{\pi \nu_\theta}{2}}\E^{i \frac{\pi}{4}}(-\conft_c)^{-\frac{3}{2}}\E^{i k \conft_c \hratio^{-1}}\left[  k \conft_c \hratio^{-1} (-2 i k \conft_c+2 i \nu_\theta +3)-2 \nu_\theta +3 i \right] \mathrm{H}_{i \nu_\theta}^{(1)}(-k \conft_c)
    \end{aligned}
\end{equation}
and
\begin{equation}
    \begin{aligned}
        B_k &= -\frac{i\pi^{\frac{1}{2}}\hratio^{2}}{4 f_a k^2}H_1 \E^{-\frac{\pi \nu_\theta}{2}}\E^{i \frac{\pi}{4}}(-\conft_c)^{-\frac{1}{2}}\E^{-i k \conft_c \hratio^{-1}}(1+ i k \conft_c \hratio^{-1}) \mathrm{H}_{1+ i \nu_\theta}^{(1)}(-k \conft_c) \\
        &-\frac{\pi^{\frac{1}{2}}\hratio^{2}}{8 f_a k^3}H_1 \E^{-\frac{\pi \nu_\theta}{2}}\E^{i \frac{\pi}{4}}(-\conft_c)^{-\frac{3}{2}}\E^{-i k \conft_c \hratio^{-1}}\left[  k \conft_c \hratio^{-1} (2 i k \conft_c+2 i \nu_\theta +3)+2 \nu_\theta -3 i \right] \mathrm{H}_{i \nu_\theta}^{(1)}(-k \conft_c)\,.
    \end{aligned}
\end{equation}
The overall factor of $\E^{-\frac{\pi}{2}\nu_\theta}$  exponential suppression is misleading as it is canceled off by the Hankel function at large $\nu_\theta$ (see footnote~\ref{fn:Boltzmann}). Note that the coefficients have acquired an extra linear dependence on $\nu_\theta$, which corresponds to an enhancement of the amplitude due to the mass. This is because while the field is $1/\nu_\theta$ suppressed, the momentum needs to be enhanced by a factor of $\nu_\theta$ in order to maintain the canonical normalization of the field. This is ensured by the Wronskian condition which requires the combination $\delta\theta_k^*  \delta\theta_k'$ to be independent of the mass of the field. This can be seen explicitly via the massive mode function. We can compute the product of the massive axion and its derivative at late times
\begin{equation}
    \begin{aligned}
       f_a^2 \delta\theta_k(\conft)\partial_\conft\delta\theta_k^*(\conft) &\xrightarrow{\conft \to 0}\frac{i}{2 a^2(\conft)}\left[1+i \frac{3}{2\nu_\theta}\coth(\pi \nu_\theta)\right] + \cdots
    \end{aligned}
\end{equation}
which ensures that
\begin{equation}
    f_a^2a^2(\conft)\delta\theta_k(\conft)\partial_\conft\delta\theta_k^*(\conft) - {\rm c.c.} = i.
\end{equation}
At late times, i.e., $\conft \to 0^{-}$, the mode functions simplify considerably
\begin{equation}
    \delta\theta_k(0^-) = A_k + B_k.
\end{equation}
And therefore the late-time power spectrum is
\begin{equation}
    P_\theta(k) = |A_k + B_k|^2\,.
\end{equation}
In terms of the scale $k_c$ defined in~\eqref{eq:defkc}, the axion power spectrum is then
\begin{equation}
    P_\theta(k) = \frac{\E^{-\pi \nu_\theta}\pi \hratio^2}{16 (f_a/H_1)^2 k^6 k_c} \left| \alpha(k) \mathrm{H}_{i \nu_\theta }^{(1)}\left(\frac{k}{k_c}\right) + \beta(k) \mathrm{H}_{i \nu_\theta +1}^{(1)}\left(\frac{k}{k_c}\right) \right|^2
\end{equation}
where we have defined the following functions for convenience
\begin{equation}\label{eq:iso_pk_funcs}
    \begin{aligned}
        \alpha(k) & = \left[2 k^2+k_c^2 (-3-2 i \nu_\theta ) \hratio\right] \sin \left(\frac{k}{k_c \hratio}\right)+k k_c (3+2 i \nu_\theta ) \cos \left(\frac{k}{k_c \hratio}\right)\,, \\
        \beta(k) &= 2 k\left[k_c \hratio \sin \left(\frac{k}{k_c \hratio}\right)-k \cos \left(\frac{k}{k_c \hratio}\right)\right]\,.
    \end{aligned}
\end{equation}
We plot this power spectrum (multiplied by $k^3$) for various choices of parameters in Fig.~\ref{fig:axion_pk}. We see that on large scales $k \ll k_c$, the power spectrum flattens out and exhibits a $k^0$ scaling whereas on small scales we recover the standard scale invariant spectrum $P_\theta \propto k^{-3}$, and will shortly analytically derive these two power laws.

We can see that, apart from changing the overall amplitude $\propto H_I/f_a$, we also have the option of lowering the amplitude by increasing $k_c$, which is essentially because we are allowing the axion mode function to dilute for a longer time while it is in its massive phase. We also observe that for our purposes changing $\hratio$ has very little effect since we are working with the assumption that $\hratio \approx 1$. We can also clearly see that increasing $\nu_\theta$ has the somewhat counterintuitive effect of raising the large scale amplitude.

For practical purposes, it is sufficient to determine the deep infrared and deep ultraviolet asymptotics of the axion power spectrum and glue the two together. As we have just seen, the power spectrum settles to a constant on large scales, whereas on short scales we return to the standard $1/k^3$ behavior of a massless scalar field. Let us now analytically estimate the power spectrum in these two regimes.

In the $k \to 0$ regime the power spectrum can be straightforwardly estimated using the small argument expansion of the Hankel functions.\footnote{In order to expand the complex conjugates, note that $\left[\mathrm{H}_{-1+i\nu}^{(1)}(z)\right]^* = -\E^{\pi \nu} \mathrm{H}_{1+i \nu}^{(2)}(z)$ and $\left[\mathrm{H}_{1+i\nu}^{(1)}(z)\right]^* = -\E^{\pi \nu} \mathrm{H}_{-1+i \nu}^{(2)}(z)$.} We find
\begin{equation}
    \begin{aligned}
        P_\theta(k) &= \frac{1}{72 \hratio f_a^2 \nu_\theta H_1}(-H_1 \conft_c)^{3}\coth(\pi \nu_\theta)\left[9+4\nu_\theta^2-36\hratio(1-\hratio)\right] \\
        &-\frac{\Gamma^2(-i \nu_\theta)}{144\pi H_1}(-H_1 \conft_c)^{3}\left(-\frac{k \conft_c}{2}\right)^{2i \nu_\theta} - {\rm c.c.} + \cdots
    \end{aligned}
\end{equation}
In the large mass limit, $\Gamma^2(\pm i \nu_\theta) \sim \E^{-\pi \nu_\theta}$ so the latter two terms, which correspond to the amplitude of particle production during inflation, are exponentially unimportant due to Boltzmann suppression. The first term, i.e.~the $k$-independent contact term is the only relevant one and is actually \textit{enhanced} in the mass. The fact that the axion power spectrum is not Boltzmann suppressed is not a surprise: the isocurvature strictly only measures the width of the $\delta\theta_k$ wavefunction, which is dominated by the so-called contact term for heavy scalars. 

The large scale power spectrum therefore simplifies to a constant
\begin{equation}
    P_\theta(k) \approx \frac{\nu_\theta}{18 H_1\hratio^2 f_a^2}(-H_1 \conft_c)^{3}\,.
\end{equation}
The exponential dilution in time is expected: a heavy field during inflation is diluted by a $1/a^3(t)$ factor due to the expansion of space. As previously discussed, the enhancement of the amplitude through the mass $\nu_\theta$ is due to the fact that the canonical momentum of the axion must be enhanced in mass even though the mode function is not.

On short scales the power spectrum simplifies to the usual scale-invariant expression for a massless scalar
\begin{equation}
    P_\theta(k) \approx \frac{H_2^2}{2 f_a^2 k^3}\,.
\end{equation}
This result matches our expectations because a scale $k$ will ``learn'' that it is massive around $\conft \approx - \nu_\theta/k$, i.e., much later for shorter scales. In other words, short-scale modes will not have enough time to learn that they are massive and will therefore be unaffected by the phase transition, resulting in the standard scale-invariant power spectrum.

Next we should estimate the scale at which the power spectrum shifts from a $k^{0}$ scaling to a $k^{-3}$ scaling within this parametrization. We can match the two asymptotic expressions to obtain
\begin{equation}
    k_* \approx k_c \hratio (\nu_\theta/9)^{-\frac{1}{3}}.
\end{equation}  
Therefore we will adopt a two-parameter broken power law power spectrum for the axion which is constant on large scales and scale-invariant on short scales
\begin{equation}
    \label{eq:P_axion}
    P_\theta(k) = \begin{cases}
        A_\theta\frac{1}{k_*^3}\,,\quad k<k_* \\
        A_\theta \frac{1}{k^3}\,, \quad k>k_*
    \end{cases}
\end{equation}
where we have defined the overall amplitude $A_\theta \equiv \left(\frac{H_2^2}{2f_a^2}\right)$. The isocurvature seen on small scales will be the standard axion isocurvature, but on large scales it is constant. Precisely such a functional form was recently studied in~\cite{Buckley:2024nen}. Our assumptions about the time at which this phase transition occurs is therefore hidden in a choice for the pivot scale $k_*$.

Finally, note that in the broken power law parametrization, the pivot scale $k_*$ can be very different from the scale we had expected to find based on general grounds at the start of this section, which was $k_t \approx \nu_\theta k_c$. This is because the role played by $k_*$ is simply to match the two asymptotic forms. Keeping all other parameters fixed, increasing $\nu_\theta$ has the effect of raising the large scale amplitude of $P_\theta(k)$ while leaving the short scale amplitude unaffected, which can be seen in Figure~\ref{fig:axion_pk}. Therefore to match the two, one must go to a $k$ smaller than $k_t$ or even $k_c$, and therefore $k_*$ scales inversely with $\nu_\theta$, albeit mildly.

\begin{figure}
    \centering
    \includegraphics[width=0.7\textwidth]{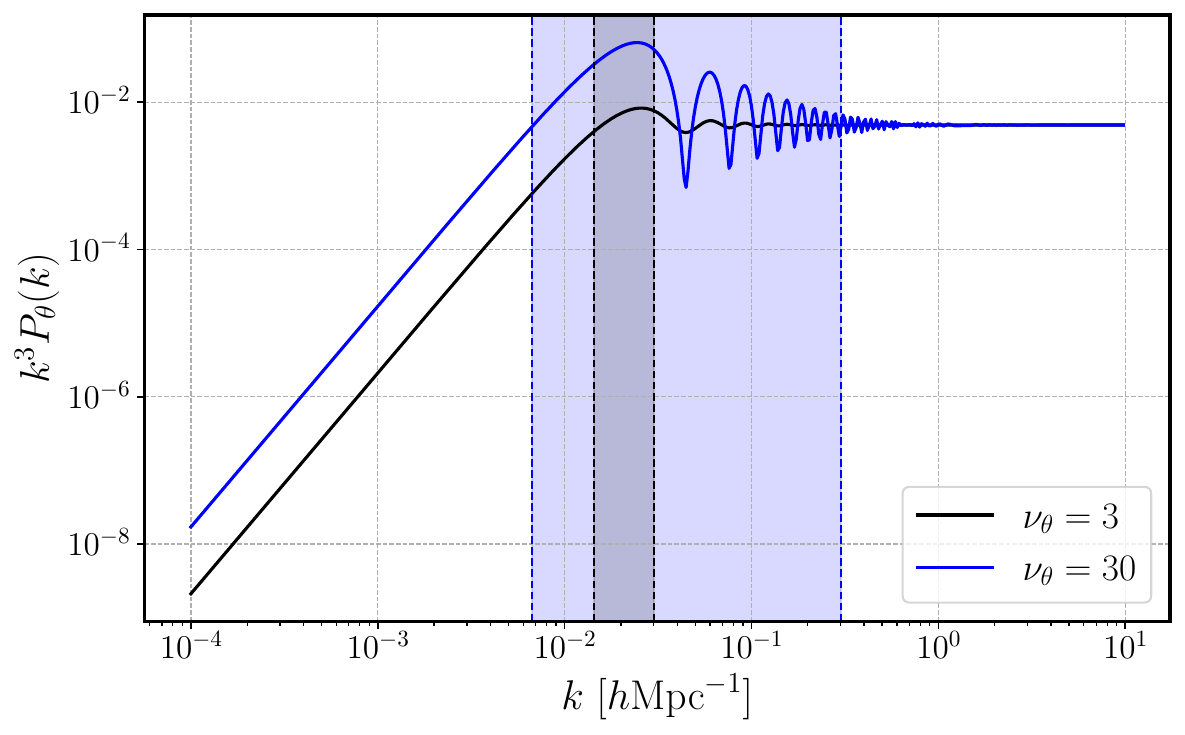}
    \caption{We plot $k^3 P_\theta(k)$ for two choices of $\nu_\theta=3$ in black and $\nu_\theta=30$ in blue. The translucent bands for each highlight the range from the pivot scale $k_* \approx k_c (\nu_\theta/9)^{-\frac{1}{3}}$ to $k_t = \nu_\theta k_c$. We clearly see that for a large axion mass, this gap spans an order of magnitude in $k$, which means that the broken power law parametrization is only appropriate for order one masses in Hubble units.}
    \label{fig:axion_pk_scales}
\end{figure}

In reality, $P_\theta(k)$ oscillates between $k_c$ and $k_t$, and subsequently settles into the scale-invariant power law, which the broken power law misses. These oscillations have a frequency set by $k_c$ and amplitude which is set by the canonical momentum $\delta\theta_k'$ and is therefore enhanced in the mass $\nu_\theta$, as is clear from Figure~\ref{fig:axion_pk_scales}. We observe these oscillations because the phase transition excites the negative frequency mode during the massless phase, and the relative size of the positive and negative frequency modes is controlled by the mass $\nu_\theta$. These features can be clearly seen from the factors $\alpha(k)$ and $\beta(k)$ in (\ref{eq:iso_pk_funcs}) which appear in the axion power spectrum. These oscillations are therefore distinct from the standard ``clock'' signal of a massive field (e.g.~\cite{Chen:2009zp}) for which the frequency and amplitude both are set by the mass, and takes the functional form $\sin(\nu_\theta \log(k))$. This signal is also present in our axion power spectrum on large scales, but as is standard, it is Boltzmann suppressed.

For small $\nu_\theta$, there are only a few modes which observe these oscillations and the power spectrum is largely controlled by the combination $k_* \propto k_c \nu_\theta^{-\frac{1}{3}}$. It is therefore not possible to disentangle the mass and critical scale for small axion masses. For the purposes of this work we will assume $\nu_\theta$ is $\mathcal{O}(1)$, such that the theory error in using the broken power law parameterization is limited to a few modes. Conversely for large masses, there are a significant number of oscillating modes between $k_c$ and $k_t$ and it is therefore possible to break this degeneracy. In that case one ought to use the full power spectrum in order to compare with data, especially since we expect the oscillations to be visible at high $k$ where more data is available. 

After the end of inflation, the axion misaligns off of the minimum of its potential by starting at some initial angle $\theta_i$, and subsequently starts to behave like dark matter. In this scenario, the inflationary fluctuations of the axion cause the dark matter and thermally generated photon fluctuations to be out of phase, and are therefore directly responsible for generating isocurvature at large scales
\begin{equation}
    \mathcal{I}_k \equiv 2\left(\frac{\Omega_a}{\Omega_c}\right) \frac{\delta\theta_k(0)}{\theta_i}\,,
\end{equation}
where $\delta\theta_k(0)$ is the axion mode function at the end of inflation. Our model predicts that this isocurvature will admit a broken power law power spectrum. Leveraging the \textit{Planck} CMB constraints on uncorrelated dark matter isocurvature, we put constraints on the $(H_I,f_a)$ parameter space in Figures~\ref{fig:axionisoconstraintsfrac},  \ref{fig:axionisoconstraintanthrop}, and~\ref{fig:axionisoconstraintdil}.

In order to implement these constraints we have assumed that the inflationary scalar power spectrum takes the standard scale-invariant form within the CMB window. We justify this choice in appendix~\ref{app:inflaton_fluct}.

%%%%%%%%%%%%%%%%%%%%%%%%%%%%%%%%%%%%%%%%%%%%		
\section{Constraints for $(H_I,f_a)$}
\label{sec:exclusionHf}
\begin{figure}
\centering
\includegraphics[]{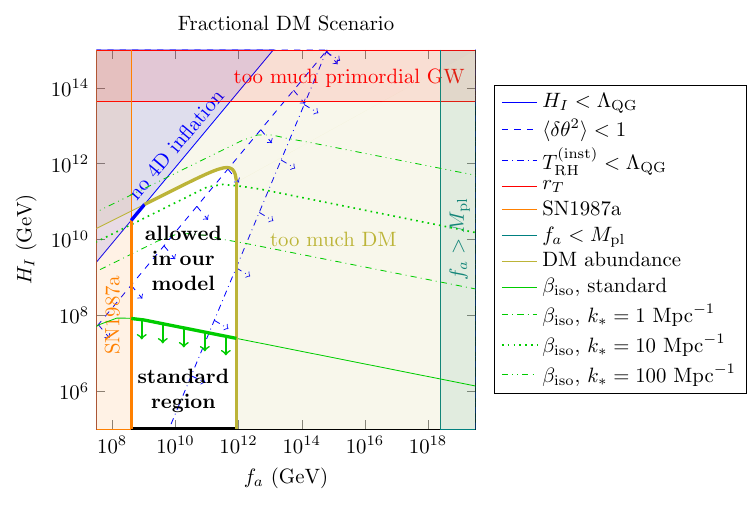}
\caption{Fractional DM scenario. Constraints on $H_I$ vs $f_a$ with $\theta_i=1$ for the standard case and our mechanism with different values of $k_*$. Shaded regions are excluded. The isocurvature constraints in the fractional DM scenario (green lines) saturate $\beta_{\text{iso}}$ by tuning $\Omega_a/\Omega_c$.}
\label{fig:axionisoconstraintsfrac}
\end{figure}

\begin{figure}
\centering
\includegraphics[]{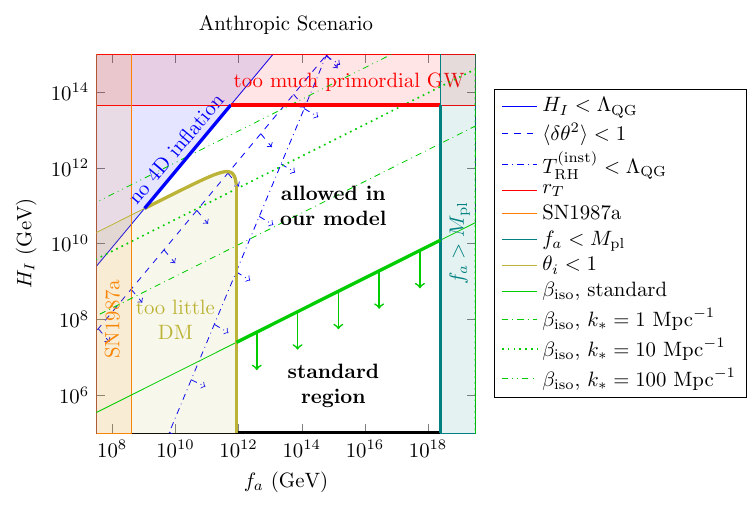}
\caption{Anthropic scenario. Constraints on $H_I$ vs $f_a$ for the standard case and our mechanism with different values of $k_*$. Shaded regions are excluded. The isocurvature constraints in the anthropic scenario (green lines) saturate both $\beta_{\text{iso}}$ and $\Omega_a$ by tuning $\theta_i$.}
\label{fig:axionisoconstraintanthrop}
\end{figure}

\begin{figure}
\centering
\includegraphics[]{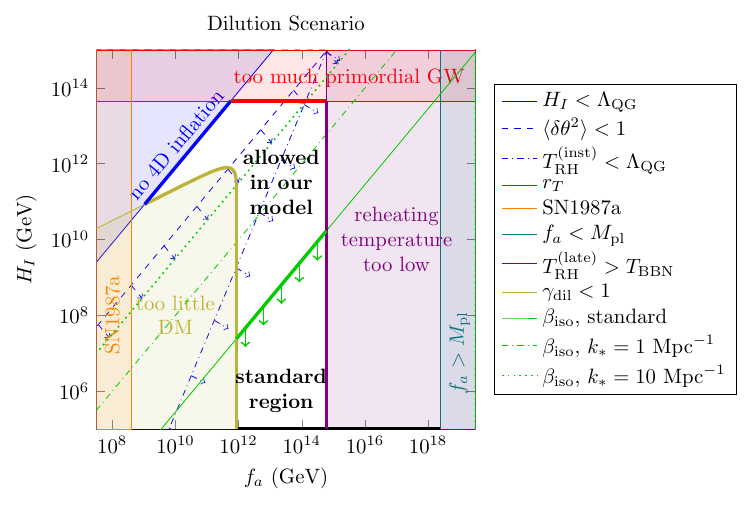}
\caption{Dilution scenario. Constraints on $H_I$ vs $f_a$ with $\theta_i=1$ for the standard case and our mechanism with different values of $k_*$. Shaded regions are excluded. The isocurvature constraints in the dilution scenario (green lines) saturate both $\beta_{\text{iso}}$ and $\Omega_a$ by tuning $\gamma_{\text{dil}}$.}
\label{fig:axionisoconstraintdil}
\end{figure}

In this section, we explain various constraints on the $(H_I,f_a)$ parameter space for three different scenarios (fractional DM, anthropic, and dilution) plotted  in Figures \ref{fig:axionisoconstraintsfrac}, \ref{fig:axionisoconstraintanthrop}, and \ref{fig:axionisoconstraintdil}, respectively. In each plot, the white region is parameter space that is allowed by our mechanism (at least potentially, up to model-dependent details), while the portion of the white region below the thick green line is allowed in the standard version of the mechanism with unsuppressed axion isocurvature.

We plot the constraint $H_I<\Lambda_{\mathrm{QG}}$ \eqref{eq:string_tension_bound} using the solid blue line. The validity of our 4D EFT also requires that $H_I<M_{\mathrm{KK}}$, where $M_{\mathrm{KK}}$ is given by \eqref{eq:M_KK} and \eqref{eq:M_s}. This is typically more restrictive than $H_I<\Lambda_{\mathrm{QG}}$, since $M_{\mathrm{KK}}$ generally lies below $\Lambda_\mathrm{QG}$. However, unlike $\Lambda_\mathrm{QG}$, $M_{\mathrm{KK}}$ has a more model-dependent dependence on $f_a$, which is controlled by gauge couplings, geometric volumes, etc. Therefore, we do not include this constraint in the plots.
Normally, one expects a homogeneous axion field in the pre-inflationary scenario. However, with small $f_a$ and large $H_I$, $\langle\delta\theta^2\rangle$ may dominate over $\theta_i^2$ and cause large random fluctuations of the axion field across superhorizon scales. This possibility has been explored for both 4D QCD axions~\cite{PhysRevD.45.3394,Linde:1990yj,Nagasawa:1991zr,Khlopov:2004sc,Kawasaki:2013iha,Co:2020dya,Liu:2020mru} and ALPs~\cite{Sakharov:2021dim,Gonzalez:2022mcx,Kitajima:2023kzu}. After the QCD phase transition, different regions may settle into different sides of the potential (e.g., at $\theta=0$ or $\theta=2\pi$), leading to the formation of axion domain walls. These domain walls are stable even for $N_{\text{DW}}=1$ (since no axion strings are formed in this scenario), and overclose the universe. We therefore impose the restriction $\langle\delta\theta^2\rangle\lesssim1$, which is plotted as the dashed blue line. This constraint applies to the standard scenario with unsuppressed axion isocurvature, or to our scenario with tunneling before the end of inflation. It does not apply to the case where tunneling ends inflation or occurs after inflation, which suppresses the isocurvature perturbations entirely.
Assuming efficient reheating after inflation, the requirement that the reheating temperature be below the 4D quantum gravity cutoff \eqref{eq:string_tension_bound} gives
\begin{gather}
    T_{\text{RH}}\sim\sqrt{H_IM_\mathrm{Pl}}<\Lambda_{\text{QG}}\lesssim2\pi\sqrt{S_{\text{inst}}}f_a,
\end{gather}
which is plotted as the blue dash-dotted line. Note that this is an aggressive bound, since inefficient reheating lowers $T_{\text{RH}}$. As explained above, we do not include the more stringent restriction $T_{\text{RH}}<M_{\mathrm{KK}}$ since it is very model-dependent.

The red line shows the upper bound on $H_I$ obtained from $r$ \eqref{eq:HItensor}.
The observation of a neutrino pulse from SN1987a places a lower bound of $f_a \gtrsim 4 \times 10^8~\mathrm{GeV}$~\cite{Raffelt:1987yt,Caputo:2024oqc}, which is plotted in orange. 
We also require $f_a<M_{\mathrm{Pl}}$, as shown in the teal line. This is an expected constraint on valid quantum gravity theories~\cite{Banks:2003sx}.\footnote{One actually expects a stronger bound, $f_a \lesssim M_\mathrm{Pl}/S_\mathrm{inst}$~\cite{Arkani-Hamed:2006emk} (the axion Weak Gravity Conjecture), but the precise numerical coefficient is somewhat uncertain~\cite{Harlow:2022ich} and may depend on the details of mixing of the QCD axion with other axions~\cite{Benabou:2025kgx}.}
For the fractional DM scenario, the yellow curve represents the constraint $\Omega_ah^2<\Omega_ch^2=0.12$. The axion dark matter abundance is given by Eq.~\eqref{eq:axiondmabundance}, where
\begin{gather}
    \langle\theta_i^2\rangle=\theta_i^2+\langle\delta\theta^2\rangle.
\end{gather}
We adopt
\begin{gather}
\langle\delta\theta^2\rangle\approx\left(\frac{4H_I}{2\pi f_a}\right)^2
\end{gather}
for the standard case \cite{PhysRevD.45.3394} and
\begin{gather}
    \langle\delta\theta^2\rangle=\int\frac{d^3k}{(2\pi)^3k^3}P_\theta(k)=\frac{H_2^2}{4\pi^2 f_a^2}\left[\frac{1}{3}+\ln\left(\frac{k_{\max}}{k_*}\right)\right]
\end{gather}
for our mechanism with tunneling during inflation. As in Section \ref{subsec:isocurvconstraint}, we set $\gamma_\mathrm{dil}=1$ and $\gamma_\mathrm{anh}=1$ if not otherwise noted. (The regions of parameter space in which anharmonic corrections are important are mostly excluded already by other considerations, so setting $\gamma_\mathrm{anh}=1$ is not unreasonable.) Because of the presence of $\langle\delta\theta^2\rangle$ in $\Omega_a$, this constraint is slightly different for the standard case and for our mechanism with different $k_*$. But this difference is almost not visible on the plot, so we only show the constraint for the standard case.

For isocurvature constraints, we use Eq.~\eqref{eq:axionisocurv} with $f_a=f_I$. The standard constraint \eqref{eq:isocurvconstraint} from Planck assumes that the isocurvature power spectrum follows a uniform functional form across all scales,
\begin{gather}
    {\cal P}_{{\cal I}{\cal I}}(k) = A_I \left(\frac{k}{k_{\text{pivot}}}\right)^{n_I - 1},
\end{gather}
and is obtained at $k_{\text{pivot}}=0.05\text{ Mpc}^{-1}$. Our mechanism predicts a different power spectrum \eqref{eq:P_axion} which critically depends on $k_*$. Since this is the same as the broken power law case studied in \cite{Buckley:2025zgh}, we can use their limits on $(\Omega_a/\Omega_c)^2A_I$ to bound $\beta_\text{iso}(k_*)$. Here we choose several different values of $k_*$ and plot the curves that saturate $\beta_\text{iso}(k_*)$ in green, with different line styles.

In the fractional DM scenario (Figure \ref{fig:axionisoconstraintsfrac}), we fix a value for $\theta_i$ so that axion may only make up a portion of dark matter. In the anthropic scenario (Figure \ref{fig:axionisoconstraintanthrop}), we assume that $\Omega_a=\Omega_c$ and use~\eqref{eq:axiondmabundance} to solve for $\theta_i$. Despite the name, only the region with large $f_a$ actually needs a fine-tuned initial misalignment angle $\theta_i\ll1$, and $\theta_i$ increases with the decrease of $f_a$. We plot in yellow the regions satisfying $\theta_i<1$, outside of which the quadratic approximation to axion potential starts to break down. The combined constraints leave open a region complementary to the allowed parameter space in the fractional DM scenario. In the dilution scenario (Figure \ref{fig:axionisoconstraintdil}), we also assume that $\Omega_a=\Omega_c$, but fix $\theta_i$ and tune $\gamma_\mathrm{dil}$ instead. As in the anthropic scenario, only the part of parameter space with large $H_I$ or $f_a$ requires fine-tuning $\gamma_\mathrm{dil}\ll1$. We exclude the regions with $\gamma_\mathrm{dil}>1$ in yellow, and the allowed parameter space typically has $f_a\gtrsim10^{12}$ GeV. For large enough values of $k_*$, the $\beta_\text{iso}(k_*)$ constraints become so loose that they can never be saturated. The dilution scenario also requires the reheating temperature to be above the BBN temperature in order to preserve the predictions of standard BBN. Specifically, we require $T_\mathrm{RH}^\mathrm{(late)} \gtrsim 6\,\mathrm{MeV}$~\cite{Barbieri:2025moq}. This, combined with $\Omega_a=\Omega_c$, sets an upper bound on $f_a$ of \cite{Kawasaki:1995vt}
\begin{gather}
    f_a\lesssim\theta_i^{-1}\,6\times10^{14}\text{ GeV}.
\end{gather}
We plot this bound in the solid violet line.

In the plots, the shaded regions are excluded by the most conservative constraints and the allowed (white) region is bounded by thick lines. The non-shaded constraints have arrows indicating the allowed region. These are either more aggressive or require additional assumptions.
A segment of the isocurvature constraint for the standard misalignment scenario separates the allowed parameter space in the standard scenario (lower region) and the parameter space opened up by our mechanism (upper region).
In all three scenarios for isocurvature constraints, the allowed parameter space is larger in our mechanism, especially with larger values of $k_*$ which correspond to later phase transition times. At some point, the isocurvature constraint becomes less restrictive than the other constraints and can thus be ignored. These exclusion plots demonstrate that our mechanism can open up parameter space for high-scale inflation, while remaining consistent with constraints from isocurvature perturbations. The same conclusion holds if the phase transition in our mechanism occurs at the end of inflation. This roughly corresponds to the limit of large $k_*$ in the plots. In this case, since the axion remains heavy throughout inflation, the isocurvature perturbations that it generates are exponentially suppressed and leave no observable imprint. The allowed parameter space is determined solely by non-isocurvature constraints.

%%%%%%%%%%%%%%%%%%%%%%%%%%%%%%%%%%%%%%%%%%%%		
\section{Toward Models and UV Completions: Extra Dimensions and String Theory}
\label{sec:models}

In this section, we discuss the embedding of the mechanism for dynamical axion monodromy mass introduced in \S\ref{subsec:dynamicalinteger} within UV completions. We focus on the case of extra-dimensional axions, where the dynamical integer $n$ can arise as a flux of a gauge field through extra dimensions. We emphasize that this section is simply a sketch of how the relevant ingredients could arise; searching for a QFT with a full higher-dimensional geometry, let alone a string theory compactification, that gives rise to the correct tunneling rate in detail is a task that we leave for future work.

\subsection{Dynamical integer from extra-dimensional flux}

We begin with a Chern-Simons term in a $d$-dimensional theory, with the $(d-4)$ extra dimensions compactified on a manifold $Y$. This theory contains a Chern-Simons term,
\begin{equation} \label{eq:xdimCSterm}
\frac{\ell}{4\pi^2} \int C_p \wedge \rmd \widetilde{B}_q \wedge \rmd A_r,
\end{equation}
with $p + q + r = d - 2$ and $\ell \in \ZZ$. (It is possible that some of the fields $A$, $\widetilde{B}$, and $C$ are the same, which would change the normalization of the prefactor.) The 4d fields are related to the $d$-dimensional fields via: 
\begin{itemize}
\item $\theta$ is the 4d axion arising from reducing $C_p$ on a $p$-cycle $\Sigma_p$, 
\begin{equation}
    \theta = \int_{\Sigma_p} C_p;
\end{equation}
\item the QCD gauge fields live on branes that also wrap the cycle $\Sigma_p$;
\item the $\widetilde{B}$ field has a flux $n_B \in \ZZ$ over a $(q+1)$-cycle $\Lambda_{q+1}$,
\begin{equation}
    n_B = \frac{1}{2\pi} \int \rmd \widetilde{B}_q \neq 0; \quad \text{and}
\end{equation} 
\item $A_3$ is a 4d zero mode from reducing $A_r$ along an $(r-3)$-cycle $\Gamma_{r-3}$,
\begin{equation}
    A_3 = \int_{\Gamma_{r-3}} A_r.
\end{equation}
\end{itemize}
We further require that these cycles can lead to the desired term under dimensional reduction, i.e., we need
\begin{equation}
    \int_Y \delta(\Sigma_p) \wedge \delta(\Lambda_{q+1}) \wedge \delta(\Gamma_{r-3}) = p \neq 0.
\end{equation}
In this case, the Chern-Simons term~\eqref{eq:xdimCSterm} reduces to the 4d term
\begin{equation}
    \frac{\ell p n_B}{2\pi} \int \theta F_4.
\end{equation}
The integer $n = \ell p n_B$ can then change dynamically via a change in the flux quantum $n_B$ through the extra dimensions.

The object magnetically charged under $\widetilde{B}_q$ is a $(d - q - 3)$-brane, or equivalently a $(p+r-1)$-brane. If we wrap this on the $p$-cycle and the $(r-3)$-cycle we are left with a 2-brane in 4 dimensions, i.e., a domain wall. This is the object across which the flux of $\rmd \widetilde{B}_q$ jumps, and accordingly the integer $n$ changes. It is electrically charged under the dual $B$ of $\widetilde{B}$, which is a 3-form gauge field in four dimensions, as in~\eqref{eq:B3kinetic}. This object is the $B$-brane that must be nucleated to produce our cosmological phase transition. 

The extra-dimensional setting also provides natural candidates for inflaton fields, in the form of the moduli fields $\{\phi_i(x)\}$ that control the size and shape of the extra dimensions. In the thin brane approximation~\eqref{eq:thinwallest}, rapid tunneling favors either large vacuum energy differences $\Delta V$ or small brane tension $\sigma$. One possibility, then, is that during inflation the geometric moduli fields evolve into a region where the volume of the cycle $\Sigma_p \times \Gamma_{r-3}$ becomes relatively small, which leads to low tension of the 4d $B$-brane and relatively easy bubble nucleation.

\subsubsection{IIA setting and the cosmological emergence of chirality}
\label{subsec:IIAchiral}

In \S\ref{subsec:dynamicalinteger}, we commented that tbe $B$-brane worldvolume theory must host a localized 2-form $b$ field. We can see this localized $b$-field explicitly in a Type IIA string theory realization. We obtain the axion $\theta$ as $\int_{\Sigma_3} C_3$, which couples to gauge fields living on D6 branes wrapped on $\Sigma_3$. The 10d action contains a Chern-Simons term of the form $C_3 \wedge \rmd C_3 \wedge \rmd B_2$. The $\rmd C_3$ term directly gives us $F_4$ in 4d, and we turn on a flux of $H_3 = \rmd B_2$ on a 3-cycle $\Lambda_3$ to obtain the desired $\theta F_4$ term. Now, the object magnetically charged under $B_2$ in 10d is an NS5 brane, which indeed has a (self-dual) 2-form gauge field on its worldvolume. When we reduce to 4d, we wrap the NS5 brane on the 3-cycle on which we reduced $C_3$ to get the axion, and it becomes the domain wall of interest. 

As recently discussed in \S6.1.1 of~\cite{Reece:2025thc}, in the context of this Type IIA UV completion, the $n \neq 0$ phase where the axion has a large monodromy mass is incompatible with the chiral spectrum of the Standard Model. To obtain chiral fermions, we would like to wrap additional D6 branes on a 3-cycle intersecting $\Sigma_3$. However, the presence of $H_3$ flux obstructs this wrapping~\cite{Freed:1999vc}. It is possible, then, that the Standard Model chiral fermions exist in the $n = 0$ phase but not in the $n \neq 0$ phase~\cite{Uranga:2002vk, Behrndt:2003ih}. This leads to the interesting cosmological possibility that most of the particle content of the Standard Model emerges after the cosmological phase transition, but is not part of the spectrum during inflation.

\subsection{Alternative: Stueckelberg axion mass with dynamical charge}

Here we briefly discuss an alternative framework, which will have very similar cosmology but could arise in a different class of UV completions. Rather than the axion obtaining a mass of the form $\frac{n}{2\pi} \theta F_4$, the axion could be eaten by an ordinary gauge field $A_1$ via the Stueckelberg mechanism. In this case, the axion carries a charge $q$ such that the gauge transformation is
\begin{equation}
    A_1 \mapsto A_1 + \rmd \lambda, \quad \theta \mapsto \theta + q \lambda,\quad q \in \mathbb{Z}.
\end{equation}
Then the axion kinetic term takes the form $\frac{1}{2} f^2 |\rmd \theta - q A|^2$ and the axion becomes the longitudinal mode of a massive spin-1 field.

In this case, we could again imagine a massive axion with $q \neq 0$ during inflation, with a phase transition to a state with $q = 0$. Although we are not accustomed to thinking of charges as dynamical, this example is not conceptually different from that of \S\ref{subsec:dynamicalinteger}. Indeed, we could dualize the axion $\theta$ to a 2-form gauge field $\widetilde{C}_2$, in which case the Stueckelberg structure dualizes to the Chern-Simons term
\begin{equation}
    \int \frac{q}{2\pi} \widetilde{C}_2 \wedge \rmd A_1.
\end{equation}
Once again, $q$ can be a dynamical integer that jumps across domain walls. These are charged under a 3-form gauge field $B_3$ for which $q$ is the electric flux. In the $B_3$ duality frame, under an $A_1$ gauge transformation 
\begin{equation}
    A_1 \mapsto A_1 + \rmd \lambda, \quad B_3 \mapsto B_3 + \lambda \widetilde{C}_2.
\end{equation}
Then the domain wall coupling $\int_M B_3$ is not gauge-invariant without an anomaly inflow mechanism, which we can realize with a localized 1-form gauge field $a_1$ on the domain wall that shifts under a $C_2$ gauge transformation. The coupling $\int_M \left[B_3 - (\rmd a_1 - \widetilde{C}_2) \wedge A_1\right]$ is then gauge-invariant.

As in the case of the monodromy mass, this EFT could arise from a higher-dimensional theory, with $q$ a magnetic flux of a gauge field through extra dimensions.

Although these structures often appear in string theory compactifications, it is often the case that a second axion arises in these settings. For instance, a linear combination of a closed string axion field and an open string axion (the phase of a charged matter field) may be eaten through the Stueckelberg mechanism, while one combination remains light. The default expectation is that it is primarily the open string axion that is eaten (see, e.g., Appendix A of~\cite{Allahverdi:2014ppa}). For our mechanism, we would want to have a light axion remaining uneaten at late times, but eaten at early times. Achieving this rather than the standard Stueckelberg axion scenario would require additional, nontrivial model-building ingredients.

%%%%%%%%%%%%%%%%%%%%%%%%%%%%%%%%%%%%%%%%%%%%		
\section{Outlook}
\label{sec:outlook}

The mechanism that we have proposed in this paper raises a number of interesting questions, both about UV completions and about detailed cosmological histories that realize the mechanism. Answers to these questions could have applications beyond this mechanism, as well.

\paragraph{Tunneling from initial states evolving in time.} Our scenario requires a tunneling rate per unit volume $\Gamma < H_I^4$ during inflation, with $\Gamma > H_I^4$ at a later time so that the transition can proceed efficiently and bubbles can merge. If this transition occurs during inflation, then we have tunneling out of a slow-roll state. If the transition occurs after inflation has ended, then the inflaton may be rolling quickly. In any case, this is a tunneling problem where the initial state does not sit in a false vacuum. Such a scenario has not been thoroughly studied. The double-field inflation literature (e.g.,~\cite{Adams:1990ds}) considers tunneling of one field triggered by the rolling of a second field that is viewed a fixed background from the viewpoint of computing the bounce action. Time dependent initial states for tunneling were studied in~\cite{Brown:2016nqt} (using a Euclidean recipe not fully developed from first principles),~\cite{Draper:2023fkz} (in the limit of small time dependence, using a quantum mechanical effective theory of the bubble radius), and \cite{Steingasser:2024ikl} (quantum tunneling from excited states using a regularized real-time formalism). It would be worthwhile to analyze time-dependent tunneling rates from first principles, to place the calculation on a firmer footing.

\paragraph{Brane effective field theory for flux tunneling.} In standard tunneling calculations, the bubble wall is a finite-size object formed by a scalar field. For flux tunneling, the bubble wall is a magnetic brane, generally treated as infinitely thin. Although it may be resolvable into an object of finite thickness (a kink in a higher-dimensional brane~\cite{Sen:1999mg,Shiu:2023bay}), we do not expect that this is required to accurately estimate the tunneling rate. However, in a QFT calculation of the tunneling rate, integrating over the brane's location in the path integral, we find divergences from varying fields at the delta-function localized brane source. We could instead set up a constrained instanton calculation where we fix the brane position, vary fields to find the bounce solution, and then subsequently maximize tunneling rates over all possible brane positions. However, this obscures the usual negative mode of the bounce (because this mode is related to the brane position, which we have fixed when finding the bounce). We expect that there is a way to set up a standard bounce calculation using a renormalized effective theory of the brane that resolves UV divergences at the brane core (along lines broadly similar to~\cite{Michel:2014lva}). It would be interesting to thoroughly understand matching to an EFT in a simpler toy model, to cleanly derive the proper procedure that would clarify this calculation and the validity of different approximations.

\paragraph{Gravitational wave signals.} 
\label{para:GR-signal}

During the final stage of the phase transition, $\gamma \gg 1$. Inside a single Hubble patch multiple vacuum bubbles nucleate, expand and collide with each other, merging into a single $n=0$ vacuum area. The collision and merging process breaks the spherical symmetry of individual bubbles. It is highly non-linear and non-perturbative, and therefore partially transfers kinetic energy stored in bubble walls into gravitational waves (GWs). For general gravitational wave signals from a first-order phase transition, the power spectrum has an asymmetric
inverted V shape around a peak frequency~\cite{Kosowsky:1992vn,Huber:2008hg} (see~\cite{Ashoorioon:2015hya} in the related double-field inflation context). For generic GW sources during inflation, the spectrum shows a unique oscillation feature at the peak frequency~\cite{An:2020fff,An:2022cce}, which can be detected by future terrestrial and astrophysical gravitational wave detectors such as LISA~\cite{LISA:2017pwj}, DECIGO~\cite{Kawamura:2011zz}, TianQin~\cite{TianQin:2015yph} and Taiji~\cite{Ruan:2018tsw}. When GWs pass through the Earth, they will lead to timing residuals in the Pulsar Timing Array (PTA) data, which can serve as an indirect detection method of GWs~\cite{Kramer:2013kea,Hobbs:2009yy,Janssen:2014dka}. The bubble collision and merging during inflation can also source curvature perturbations and the proceeding secondary GWs (see~\cite{An:2023jxf}). It would be interesting to quantify the detectability of such signals for our scenario.

\paragraph{Reheating after the first-order phase transition.} As depicted in Fig.~\ref{fig:merger}, the merger of bubbles in the first-order phase transition leaves behind localized regions of $n \neq 0$ vacuum inside the low-energy $n = 0$ vacuum, bounded by $B$-brane walls. These regions collapse and decay into radiation. How quickly do they collapse? Into what degrees of freedom do they mostly decay? If the tunneling process ends inflation or occurs after inflation ends, the answers to these questions determine the physics of reheating. It would be interesting to find a setting in which the details of this reheating process are calculable.

\paragraph{Emergence of chirality after a first-order phase transition.} The Type IIA setting discussed in \S\ref{subsec:IIAchiral} raises the interesting cosmological possibility that the Standard Model, with its chiral fermions, is an emergent phenomenon arising after inflation (in bubble interiors). One potential application is to the ``Festina Lente'' conjecture that appears to rule out charged chiral fermions during inflation~\cite{Montero:2019ekk, Montero:2021otb}. If this conjecture holds up to scrutiny (e.g.,~\cite{Aalsma:2023mkz,Lin:2024jug}), it favors cosmologies with emergent late-time chirality. Emergent chirality could also have interesting implications for baryogenesis. It would be interesting to explore whether the clash between an axion monodromy mass and the existence of chiral charged fermions persists in a wider range of UV completions.

%%%%%%%%%%%%%%%%%%%%%%%%%%%%%%%%%%%%%%%%%%%%		
\section*{Acknowledgments}

We thank Itamar Allali and JiJi Fan for useful discussions about isocurvature perturbations. MR also thanks Michele Cicoli, Thomas Grimm, Jim Halverson, Fernando Marchesano, and Jake McNamara for useful questions or comments when an early version of this work was presented at the String Pheno 2024 conference, and Fernando Marchesano for further helpful correspondence. We would like to thank Rashmish Mishra and Michael Nee for useful conversations. This work is supported in part by the DOE Grant DE-SC0013607. This work was performed in part at the Aspen Center for Physics, which is supported by the National Science Foundation grant PHY-2210452.
			
\appendix

\section{Inflaton fluctuations}\label{app:inflaton_fluct}

For completeness, we will also briefly discuss the effect of this phase transition on the inflaton perturbations, specifically for the cosmological scenario studied in section~\ref{sec:casestudy}. At leading order in our EFT, the axion and the inflaton fluctuations are not coupled on large scales and therefore the comoving curvature
\begin{equation}\label{eq:comoving_curvature}
    \mathcal{R}_k \equiv -\frac{H_I}{\dot{\bar{\phi}}}\delta\phi_k 
\end{equation}
must be unaffected on scales which have exited the horizon before the phase transition, i.e.~for $k<k_c$. Therefore any effects of the phase transition must only be visible on short scales which we assume to be outside the CMB window. As such it is not essential for our purposes to be quantitatively precise and we will only proceed with some qualitative remarks about the short scale scalar power spectrum.

There are mainly two possible ways for the inflaton fluctuations, and therefore the comoving curvature perturbations, to be affected. One is through the change in $H_I$ and the slow-roll parameters $\epsilon$ and $\eta$. While we have assume $H_I$ does not change substantially through the phase transition, the same is not necessarily true for the slow-roll parameters. Since $\mathcal{R}_k \propto H_I/(\sqrt{2 \epsilon} M_{\rm pl})$ both the amplitude and the spectral index can in principle be different on short scales, depending on the precise inflaton potential after the phase transition is over.

On the other hand, the nucleated bubbles will collide and subsequently deposit energy density into the inflaton fluctuations. This would correspond to a deviation from a scale-invariant spectrum around the critical scale $k_c$ or below. Crucially, since these bubbles are horizon sized at the time of the phase transition, modes with $k< k_c$ are unaffected. 

Since the energy density stored in the bubble walls is fixed by the vacuum splitting, we can estimate the energy density sourced by the bubble collisions to be $\sim \Delta V$. This is to be weighed against the typical size of massless inflaton fluctuations and therefore the size of this effect is controlled by the ratio $\Delta V/H_I^4$ \cite{An:2023jxf}. As discussed in section~\ref{sec:cosmology}, two of our key assumptions are that the bubble nucleation rate admits a semiclassical description and that gravitational effects are subdominant. The latter requires $\varepsilon_\rho \equiv 3 \sigma H_I/\Delta V \ll 1$. To ensure the validity of the semiclassical formula (\ref{eq:thinwallest}) we need a large instanton action
\begin{equation}
    S_0 \sim \varepsilon_\rho^4 \frac{\Delta V}{H_I^4} \gg 1.
\end{equation}

Thus within our model assumptions, we expect that the ratio $\Delta V/H_I^4$ is large and therefore this effect is important. Unfortunately, a precise quantitative analysis is model dependent and can require numerical simulations, which is beyond the scope of this work. In practice, especially pertaining to the constraints shown in section~\ref{sec:exclusionHf}, we will assume $k_c \gtrsim 1 \,{\rm Mpc}^{-1}$ so that the CMB isocurvature constraints obtained in \cite{Buckley:2025zgh} can be robustly applied to our model. It is interesting to quantify the impact of such bubble collisions on both the scalar and tensor power spectrum as these would provide yet another signature of the first order phase transition (see e.g.~\cite{An:2023jxf} for an analysis in the context of tunneling in a double field inflation model). We leave this for future work.

\bibliography{ref}
\bibliographystyle{utphys}

\end{document}